\documentclass[prd,twocolumn,superscriptaddress,nofootinbib]{revtex4-1}
\usepackage{epsfig}
\usepackage{amsmath}
\usepackage{amsfonts}
\usepackage{float}
\usepackage{amssymb}   
\usepackage{slashed}
\usepackage{color} 
\usepackage{multirow}
\usepackage{pbox}   
\usepackage{subfigure}
\usepackage{array,multirow,makecell}
\usepackage[colorlinks,citecolor=blue]{hyperref}
\usepackage{tabularx} 
\usepackage{titlesec} 
\usepackage{scrextend}
\usepackage{ulem}
\usepackage{natbib}
\usepackage{array,multirow,makecell}
\usepackage{lipsum}
\usepackage[colorlinks,citecolor=blue]{hyperref}

\def\lsim{\raise0.3ex\hbox{$\;<$\kern-0.75em\raise-1.1ex\hbox{$\sim\;$}}}
\def\gsim{\raise0.3ex\hbox{$\;>$\kern-0.75em\raise-1.1ex\hbox{$\sim\;$}}}
\def\be{\begin{equation}}
\def\ee{\end{equation}}
\def\bea{\begin{eqnarray}}
\def\eea{\end{eqnarray}}

\titleformat{\paragraph}
{\normalfont\normalsize\bfseries}{\theparagraph}{1em}{}
\titlespacing*{\paragraph} 
{0pt}{3.25ex plus 1ex minus .2ex}{1.5ex plus .2ex}
\begin{document}

\title{Thermal keV dark matter in a gauged B-L model with cosmic inflation}
\author{Debasish Borah}
\email{dborah@iitg.ac.in}
\affiliation{Department of Physics, Indian Institute of Technology Guwahati, Assam 781039, India}

\author{Suruj Jyoti Das }
\email{suruj@iitg.ac.in}
\affiliation{Department of Physics, Indian Institute of Technology Guwahati, Assam 781039, India}

\author{Abhijit Kumar Saha}
\email{psaks2484@iacs.res.in}
\affiliation{School of Physical Sciences, Indian Association for the Cultivation of Science, 2A $\&$ 2B Raja S.C. Mullick Road, Kolkata 700 032, India}

%

\begin{abstract}
We investigate the possibility of keV scale thermal dark matter (DM) in a gauged $B - L$ extension of the standard model with three right-handed neutrinos (RHN) and one vector like fermion in the context of cosmic inflation. The complex singlet scalar
field responsible for the spontaneous breaking of $B-L$ gauge symmetry is non-minimally coupled to gravity and serves the role of inflaton. The keV scale vector like fermion DM gives rise to the possibility of warm dark matter, but it gets overproduced thermally. The subsequent entropy dilution due to one of the RHN decay can bring the thermal abundance of DM within the observed limit. The dynamics of both the DM and the diluter are regulated by the $B-L$ model parameters which are also restricted by the requirement of successful inflationary dynamics.
We constrain the model parameter space from the requirement of producing sufficient entropy dilution to obtain correct DM relic,  inflationary observables along with other phenomenological constraints. Interestingly, we obtain unique predictions for the order of lightest active neutrino mass ($m_{\nu_l}\lesssim 10^{-14}$ eV) as function of $B-L$ gauge coupling for keV scale DM. The proposed framework also explains the origin of observed baryon asymmetry from the decay of other two heavier RHN by overcoming the entropy dilution effect. 

\end{abstract}
\maketitle
\section{Introduction}
\label{sec:intro}
The existence of dark matter (DM) occupying a significant amount in the energy budget of the universe is firmly established after the observations of different astrophysical and cosmological experiments, as summarised in review articles \cite{Jungman:1995df,Feng:2010gw,Bertone:2004pz}. The observed anisotropies in the cosmic microwave background (CMB) measurements provide a precise estimate of the DM abundance quoted as $\Omega_{\rm DM} h^2 =0.120 \pm 0.001$ at $68\%$ CL \cite{ParticleDataGroup:2020ssz,Planck:2018vyg}, where $\Omega$ refers to density parameter and $h$ is the reduced Hubble constant in the unit of 100 km/sec/Mpc. Since none of the standard model (SM) particles has the required  
properties of a DM particle\footnote{It is worth mentioning that neutrinos in the SM satisfy some of the criteria for being a good DM candidate, but due to their sub-eV masses, they remain relativistic at the epoch of freeze-out as well as matter radiation equality, giving rise to hot dark matter (HDM) which is ruled out by both astrophysics and cosmology observations.}, construction of beyond Standard model (BSM) frameworks has become a necessity. Among different BSM frameworks for such particle DM, weakly interacting massive particle (WIMP) is the most popular one where a DM particle having mass and interactions similar to those around the electroweak scale gives rise to the observed relic after thermal freeze-out, a remarkable coincidence often referred to as the {\it WIMP Miracle} \cite{Kolb:1990vq}. Such interactions enable the WIMP DM to be produced in thermal equilibrium in the early universe and eventually its number density gets frozen out when the rate of expansion of the universe takes over the interaction rates. Such DM candidates typically remain non-relativistic at the epochs of freeze-out as well as matter-radiation equality and belong to the category of Cold Dark Matter (CDM). However, the absence of any observational hints for WIMP at direct search experiments namely LUX \cite{LUX:2016ggv}, PandaX \cite{PandaX-4T:2021bab} and Xenon1T \cite{XENON:2018voc} or collider experiments such as the large hadron collider (LHC) \cite{Kahlhoefer:2017dnp,Penning:2017tmb} has also encouraged the particle physics community to pursue other viable alternatives.      

As mentioned above, the HDM paradigm is disfavoured due to its relativistic nature giving rise to a large free
streaming length (FSL) \cite{Boyarsky:2008xj,Merle:2013wta} which can erase small-scale structures. One intermediate possibility is the so called Warm Dark Matter (WDM) scenario where DM can remain semi-relativistic during the epoch of matter radiation equality. The FSL of WDM also falls in the intermediate regime between those of CDM and HDM. Typically, WDM candidates have masses in the keV regime which once again stays intermediate between sub-eV scale masses of HDM and GeV-TeV scale masses of CDM. A popular choice of WDM candidate is a right handed sterile neutrino, singlet under the SM gauge symmetry. A comprehensive review of keV sterile neutrino DM can be found in \cite{Drewes:2016upu}.
To assure the stability of DM, the sterile neutrino should have tiny or zero mixing with the SM neutrinos leading to a lifetime larger than the age of the Universe. The lower bound on WDM mass arises from observations of the Dwarf spheroidal galaxies
and Lyman-$\alpha$ forest which is around 2 keV \cite{Gorbunov:2008ka,Boyarsky:2008ju,Boyarsky:2008xj,Seljak:2006qw}. Although WDM may not be traced at typical direct search experiments or the LHC, it can have interesting signatures at indirect search experiments. For example, a sterile neutrino WDM candidate having mass 7.1 keV can decay on cosmological scales to a photon and a SM neutrino, indicating an origin to the unidentified 3.55 keV X-ray line reported by analyses of refs. \cite{Bulbul:2014sua} and \cite{Boyarsky:2014jta} using the data collected by the XMM-Newton X-ray telescope. On the other hand, WDM can have interesting astrophysical signatures as it has the potential to solve some of the small-scale structure problems of the CDM paradigm \cite{Bullock:2017xww}. For example, due to small FSL, the CDM paradigm gives rise to formation of structures at a scale as low as that of the solar system, giving rise to tensions with astrophysical observations. The WDM paradigm, due to its intermediate FSL can bring the predictions closer to observations. 

The CMB experiments also reveal that our universe is homogeneous and isotropic
on large scales upto an impressive accuracy. Such observations lead to the so called horizon and flatness problems which remain unexplained in the description of standard cosmology. Theory of cosmic inflation which accounts for the presence of a rapid accelerated expansion phase in the early universe, was proposed in order to alleviate these problems \cite{Guth:1980zm,Starobinsky:1980te,Linde:1981mu}. The Higgs field from the SM sector \cite{Bezrukov:2007ep}, has the ability to serve the role of inflaton, but this particular description suffers from the breakdown of perturbative unitarity \cite{Lerner:2009na}. The simple alternative is to extend the SM of particle physics by a gauge singlet scalar which can be identified with the inflaton.

Motivated by the pursuit of finding a common framework for cosmic inflation and WDM, in this work, we adopt a minimally extended gauged $B-L$ scenario which includes three right handed neutrinos (RHN) required to cancel the gauge anomalies and one vector like fermion, the candidate for dark matter. The model also contains a complex singlet scalar to spontaneously break the $B-L$ gauge symmetry while simultaneously generating RHN masses. The presence of RHN naturally assist to accommodate active neutrino masses (which SM by itself can not explain) via type-I seesaw.

 The study of cosmic inflation with the scalar singlet as the inflaton in minimal $B-L$ has been performed earlier in \cite{Okada:2011en,Okada:2015lia}. Further extension of the minimal $B-L$ model
with a discrete symmetry $Z_2$ can provide a stable dark matter candidate which is the lightest RH neutrino. An attempt to simultaneous realisation of inflation and cold dark matter (both thermal and non-thermal) considering minimal $U(1)_{B-L}$ model has been performed in \cite{Borah:2020wyc}.
In this work, we envisage the scope of realizing warm dark matter in a $B-L$ inflationary framework which also offers correct magnitude of spectral index and tensor to scalar ratio in view of most recent Planck 2018+BICEP/Keck data \cite{BICEPKeck:2021gln} published in 2021. For the purpose we have added a new gauge singlet vector like fermion (serving as the DM candidate) to the minimal gauged $B-L$ model. Such extension is desirable for simultaneous explanation of correct WDM relic and active neutrino masses satisfying neutrino oscillation data as we explain in upcoming sections.
 
 The thermally produced keV scale warm dark matter in a gauged $U(1)_{B-L} $ model turns overabundant. As a remedy, the relic can be brought down to the observed limit by late time entropy production due to decay of a long lived RHN. Previously, such dilution mechanism of thermally overproduced keV scale dark matter abundance from long lived heavier RHN decay has been studied in a broader class of models having $U(1)_{B-L}$ gauge symmetry without accommodating cosmic inflation by the authors of \cite{Nemevsek:2012cd, Bezrukov:2009th, Borah:2017hgt, Dror:2020jzy,Dutra:2021lto, Arcadi:2021doo}. In particular, it has been shown earlier in \cite{Bezrukov:2009th} considering generic gauge extensions, such entropy dilution for a sterile neutrino keV scale dark matter works against the pursuit of fitting neutrino oscillation data. Here we work in a suitable extended gauged $B-L$ model that is able to address keV scale thermal dark matter, neutrino masses and mixing and the cosmic inflation at the same time. We assume $N_3$ (the heaviest RHN) to be in thermal equilibrium at the early universe and its late decay depletes the thermally overproduced keV scale DM by entropy dilution. The requirement of successful inflation provides strong bounds on additional parameters namely, the $B-L$ gauge coupling and the singlet scalar-RHN coupling and combination of them. Such constraints have important implications on thermal history of $N_3$ and subsequently on WDM phenomenology. These constraints, in general, can affect the realisation of thermalisation condition for $N_3$ (also DM) as well as the decoupling temperature of $N_3$. The decoupling temperature of $N_3$ or the (non-relativistic) freeze out abundance of $N_3$ is very sensitive to the required amount of late entropy injection which in turn, dilutes the overproduced thermal WDM relic. In addition, the mass scales of the additional gauge boson and the diluter RHN can not be arbitrary in view of constraints arising from inflationary dynamics. With these impressions, the present study is important to combine thermally produced keV scale dark matter phenomenology in presence of long-lived $N_3$ with cosmic inflation in a gauged $B-L$ model which is not performed earlier to the best of our knowledge. To mention some of the important findings of the present study, we report the surviving parameter spaces after accommodating phenomenological constraints arising from fitting dark matter relic abundance, cosmic inflation, thermalisation of dark matter and the diluter in early universe and direct searches at colliders like the LHC. Next, we demonstrate distinct predictions for the lightest active neutrino mass ($\lesssim 10^{-14}$\,eV) as function of $B-L$ model parameters considering keV scale DM mass. We also find that it is still possible to produce the observed baryon symmetry in the universe via leptogenesis even with the significant entropy dilution effect at late epochs to satisfy the DM relic. 

The paper is organized as follows. In section \ref{sec:model}, we briefly discuss the minimal gauged $B-L$ model followed by summary of scalar singlet inflation via non-minimal coupling to gravity in section \ref{sec:inflation}. In section \ref{sec:DM}, we discuss the procedure to calculate the relic density of keV RHN DM followed by discussion of DM related results incorporating other bounds. We comment on neutrino mass and leptogenesis in section\,\ref{sec:NeuMass}. Finally, we conclude in section \ref{sec:conclude}.

\section{Gauged $B-L$ Model}\label{sec:model}
As mentioned earlier, gauged $B-L$ extension of the SM \cite{Davidson:1978pm, Mohapatra:1980qe, Marshak:1979fm, Masiero:1982fi, Mohapatra:1982xz, Buchmuller:1991ce} is one of the most popular BSM frameworks. In the minimal version of this model, the SM particle content is extended with three right handed neutrinos ($N_R$) and one complex singlet scalar ($\Phi$) all of which are singlet under the SM gauge symmetry. The requirement of triangle anomaly cancellation
fixes the $B-L$ charge for each of the RHNs as -1. The complex singlet scalar having $B-L$ charge 2 not only leads to spontaneous breaking of gauge symmetry but also generate RHN masses dynamically. 
In this work, we simply add one vector like fermion $\chi$ to the minimal guaged $B-L$ model having $B-L$ charge $q_\chi \neq 0$. In order to ensure its stability we choose $q_\chi \neq 1$ and fix it at an order one value of $4/3$ for numerical calculations in our analysis.

The gauge invariant Lagrangian of the model can be written as
\begin{eqnarray}
\mathcal{L}&=&\mathcal{L}_{\rm SM} -\frac{1}{4} {B^{\prime}}_{\alpha \beta}
\,{B^{\prime}}^{\alpha \beta} + \mathcal{L}_{\rm scalar} 
+ \mathcal{L}_{\rm fermion}\;,
\label{LagT}
\end{eqnarray} 
where $\mathcal{L}_{\rm SM}$ represents the SM Lagrangian involving quarks,
gluons, charged leptons, left handed neutrinos and electroweak gauge
bosons. The second term in the $\mathcal{L}$ indicates the kinetic term of $B-L$ gauge boson ($Z_{BL}$),
expressed in terms of field strength tensor ${B^\prime}^{\alpha\beta}=
\partial^{\alpha}Z_{BL}^{\beta}-\partial^{\beta}Z_{BL}^{\alpha}$. The gauge invariant scalar Lagrangian of the model (involving SM Higgs $H_{\rm S}$ and singlet scalar $\Phi)$ is given by,
\begin{align}
 \mathcal{L}_{\rm scalar}=(D_{\mu} H_{\rm S})^\dagger (D^{\mu} H_{\rm S})+(D_{\mu} \Phi)^\dagger (D^{\mu} \Phi)-V(H_{\rm S},\Phi)~,
\end{align}
where,
\begin{align}
V(H_{\rm S},\Phi)=-\mu_{1}^{2} |H_{\rm S}|^{2}-\mu_{2}^{2}|\Phi|^{2}+\lambda_{1} |H_{\rm S}|^{4}\nonumber\\
+\lambda_{2}|\Phi|^{4}+\lambda_{3} |H_{\rm S}|^{2}|\Phi|^{2}.\label{eq:PotI}
\end{align}
The covariant derivatives of scalar fields are written as,
\begin{align}
 &D_{\mu} H_{\rm S}=\left(\partial_\mu+i\frac{g_1}{2}\sigma_aW^a_{\mu}+i\frac{g_2}{2}B_\mu\right)H_{\rm S},\\
  &D_{\mu} \Phi =\left(\partial_\mu+i2g_{\rm BL}Z_{BL\mu}\right)\Phi,
\end{align}
with $g_1$ and $g_2$ being the gauge couplings of $SU(2)_L$ and $U(1)_Y$ respectively
and $W^a_{\mu}$
($a=1,\,2,\,3$), $B_{\mu}$ are the corresponding gauge fields. On the other hand $Z_{BL}, g_{BL}$ are the gauge boson and gauge coupling respectively for $U(1)_{B-L}$ gauge group. There can also be a kinetic mixing term between $U(1)_Y$ of SM and $U(1)_{B-L}$ of the form $\frac{\epsilon}{2} B^{\alpha \beta} B^{\prime}_{\alpha \beta}$ with $\epsilon$ being the mixing parameter. While this mixing can be assumed to be vanishing at tree level, it can arise at one-loop level as $\epsilon \approx g_{\rm BL} g_2/(16 \pi^2)$ \cite{Mambrini:2011dw}. As we will see in upcoming sections, the $B-L$ gauge coupling $g_{\rm BL}$ has tight upper bound from inflationary dynamics, and hence the one-loop kinetic mixing can be neglected in comparison to other relevant couplings and processes. Therefore, for simplicity, we ignore such kinetic mixing for the rest of our analysis.  

The gauge invariant Lagrangian involving RHNs and DM $\chi$ can be written as
\begin{align}
 \mathcal{L}_{\rm fermion} &=  i \sum_{\kappa=1}^{3}\overline{N_{R_{\kappa}}} \slashed{D}(Q^R_{\kappa}) N_{R_{\kappa}} -\sum_{\substack{j=1~\\ \alpha=e, \mu, \tau}}^{3}Y_D^{\alpha j}~\overline{l_{L}^{\alpha}}\tilde{H_{\rm S}}N_{R}^{j}\nonumber\\
 &-\sum_{i,j=1}^{3}Y_{N_{ij}}\Phi~\overline{N_{R_i}^{C}}N_{R_j}+ \overline{\chi}\gamma^\mu(\partial_\mu+ig_{BL}q_\chi Z_{BL_\mu})\chi\nonumber \\&~-m_\chi\overline{\chi}\chi.
+{\rm h.c.}\label{eq:Lferm}
\end{align}
The covariant derivative for $N_{R \kappa}$ is defined as 
\begin{eqnarray}
\slashed{D}(Q^{R}_{\kappa})\,{N_{R_{\kappa}}} =
\gamma^{\mu}\left(\partial_{\mu}
+ i g_{BL}\,Q^{(R)}_{\kappa}\,{Z_{BL}}_{\mu}\right) {N_{R_{\kappa}}} \,, 
\end{eqnarray}
with $Q^R_{\kappa}=-1$ is the $B-L$ charge of right handed neutrino $N_{R_{\kappa}}$. Hereafter, we denote the RHNs by $N_i, i=1,2,3$ only without explicitly specifying their chirality. 

After spontaneous breaking of both $B-L$ symmetry and electroweak symmetry,
the SM Higgs doublet and singlet scalar fields are expressed as,
\begin{eqnarray}
H_{\rm S}=\begin{pmatrix}H_{\rm S}^+\\
\dfrac{h + v + i A}{\sqrt{2}}\end{pmatrix}\,,\,\,\,\,\,\,
\Phi = \dfrac{\phi+v_{BL}+ iA^{\prime}}{\sqrt{2}}
\label{H&phi_broken_phsae}
\end{eqnarray}
where $v$ and $v_{BL}$ are vacuum expectation values (VEVs) of $H_{\rm S}$
and $\Phi$ respectively. The right handed neutrinos and $Z_{BL}$ acquire masses after the $U(1)_{B-L}$ breaking as, 
\begin{align}
 &M_{Z_{BL}}=2 g_{BL} v_{BL,}\label{eq:Zmass}\\
 &M_{N_{i}}=\sqrt{2}Y_{N_{i}}v_{BL}.\label{eq:DMmass}
\end{align}
We have considered diagonal Yukawa matrix $Y_N$ in the $(N_{1},N_{2}, N_{3})$ basis. Using Eq.(\ref{eq:Zmass}) and Eq.(\ref{eq:DMmass}), it is possible to relate  $M_{Z_{BL}}$ and $M_{N_i}$ by,
\begin{align}
 M_{N_{i}}=\frac{1}{\sqrt{2} g_{BL}}Y_{N_{i}} M_{Z_{BL}}.
\end{align}
After the spontaneous breaking of $SU(2)_{L}\times U(1)_{Y}\times U(1)_{B-L}$ gauge symmetry, the mixing between scalar fields $h$ and $\phi$ appears and can be related to the physical mass eigenstates $H_{1}$ and $H_{2}$ by a rotation matrix as, 
\begin{align}
\begin{pmatrix}H_{1}\\
H_{2}
\end{pmatrix}=\begin{pmatrix}\cos\theta & -\sin\theta\\
\sin\theta & \cos\theta
\end{pmatrix}\begin{pmatrix}h\\
\phi
\end{pmatrix},
\end{align}
where the scalar mixing angle $\theta$ is found to be 
\begin{align}
\tan2\theta=-\frac{\lambda_{3}vv_{BL}}{(\lambda_{1}v^2-\lambda_{2}v_{BL}^2)}~.
\label{mixingangle}
\end{align}  
The mass eigenvalues of the physical scalars are given by, 
\begin{align}
 &M_{H_{1}}^2=2{\lambda_{1}}v^2\cos^2\theta+2{\lambda_{2}}v_{BL}^2\sin^2\theta-2{\lambda_{3}}vv_{BL}\sin\theta\cos\theta,\\
 &M_{H_{2}}^2=2{\lambda_{1}}v^2\sin^2\theta+2{\lambda_{2}}v_{BL}^2\cos^2\theta+2{\lambda_{3}}vv_{BL}\sin\theta\cos\theta.\label{eq:HiggsEigen}
\end{align}
Here $M_{H_1}$ is identified as the SM Higgs mass whereas $M_{H_2}$ is the singlet scalar mass.

One of the strong motivations of the minimal $U(1)_{B-L}$ model is the presence of heavy RHNs which can yield correct light neutrino mass via type I seesaw mechanism. The analytical expression for the light neutrino mass matrix is
\begin{align}
 m_{\nu}=m_D M_{N}^{-1}m_D^T,
\end{align}
where $m_D=Y_Dv/\sqrt{2}$. We consider the right handed neutrino mass matrix $M_N$ to be diagonal. Since in our case one of the RHNs (say $N_{3}$) is long lived, it  interacts with SM leptons very feebly, the lightest active neutrino would be very tiny.

There are several theoretical and experimental constraints that restrict the model parameters of the minimal $U(1)_{B-L}$ model. To begin with, the criteria to ensure the scalar potential bounded from below yields following conditions involving the quartic couplings,
\begin{align}
\lambda_{1, 2, 3} \geq 0, \, \lambda_3 + \sqrt{\lambda_1 \lambda_2} \geq 0
\end{align}
\noindent On the other hand, to avoid perturbative breakdown of the model, all dimensionless couplings must obey the following limits at any energy scale,
\begin{align}
\lambda_{1,2,3} | < 4\pi, \, |Y_D, Y_N| < \sqrt{4\pi}, \, |g_1, g_2, g_{BL} | < \sqrt{4\pi}.
\end{align}

The non-observation of the extra neutral gauge boson in the LEP experiment \cite{Carena:2004xs,Cacciapaglia:2006pk} imposes the following constraint on the ratio of $M_{Z_{BL}}$ and $g_{BL}$ :
\begin{align}
 \frac{M_{Z_{BL}}}{g_{BL}} \geq 7 {~\rm TeV}.
\end{align}
 The recent bounds from the ATLAS experiment \cite{Aad:2019fac} and the CMS experiment \cite{Sirunyan:2018exx} at the LHC rule out additional gauge boson masses below 4-5 TeV from analysis of 13 TeV centre of mass energy data. However, such limits are derived by considering the corresponding gauge coupling $g_{BL}$ to be similar to the ones in electroweak theory and hence the bounds become less stringent for smaller gauge couplings.

Additionally, the parameters associated with the singlet scalar of the model are also constrained \cite{Robens:2015gla,Chalons:2016jeu} due to the non-zero scalar-SM Higgs mixing. The bounds on scalar  singlet-SM Higgs mixing angle arise from several factors namely $W$ boson mass correction \cite{Lopez-Val:2014jva} at NLO, requirement of perturbativity and unitarity of the theory, the LHC and LEP direct search \cite{Khachatryan:2015cwa,Strassler:2006ri} and Higgs signal strength measurement \cite{Strassler:2006ri}. If the singlet scalar turns lighter than SM Higgs mass, SM Higgs can decay into a pair of singlet scalars. Latest measurements by the ATLAS collaboration restrict such SM Higgs decay branching ratio into invisible particles to be below $13\%$ \cite{ATLAS:2020cjb} at $95\%$ CL. 

In our case, we work with very small singlet scalar-SM Higgs mixing and considered all the scalar quartic couplings to be positive. These help in evading the bounds on scalar singlet mixing angle and the boundedness of the scalar potential from below.
We choose the magnitude of the relevant couplings below their respective pertubativity limits. 

\section{Cosmic Inflation}
\label{sec:inflation}
 Here we briefly discuss the dynamics of inflation in view of most recent data from combination of Planck and BICEP/Keck \cite{BICEPKeck:2021gln}. For a detailed discussion of inflation in the minimal B-L model, we refer to \cite{Okada:2011en, Okada:2015lia}. We identify the real part ($\phi$) of singlet scalar field $\Phi$ as the inflaton. Along with the renormalisable potential in Eq.(\ref{eq:PotI}), we also assume the presence of non-minimal coupling of $\Phi$ to gravity. The potential that governs the inflation is given by
\begin{align}
 V_{\rm Inf}(\phi)=\frac{\lambda_2}{4} \phi^4+ \frac{\xi}{2} \phi^2 R,\label{eq:PotPhi}
\end{align}
where $R$ represents the Ricci scalar and $\xi$ is a dimensionless coupling of singlet scalar to gravity. We have neglected the contribution of $v_{BL}$ in Eq.(\ref{eq:PotPhi}) by considering it to be much lower than the Planck mass scale $(M_P)$. With this form of potential, the action for $\phi$ in Jordan frame is expressed as,
\begin{align}
 S_J=\int d^4x\sqrt{-g}\Bigg[-\frac{M_P^2}{2}\Omega(\phi)^2R+\frac{1}{2}(D_{\mu}\phi)^\dagger(D^{\mu}\phi)-\frac{\lambda_2}{4} \phi^4\Bigg],
\end{align}
where $\Omega(\phi)^2=1+ \frac{\xi\phi^2}{M_P^2}$, $g$ is the spacetime metric in the $(-,+,+,+)$ convention, $D_{\mu} \phi$ stands for the covariant derivative of $\phi$ containing couplings with the gauge bosons which reduces to the normal derivative $D_\mu\rightarrow \partial_\mu$ (since during inflation, the SM and BSM fields except the inflaton are non-dynamical).

\begin{figure*}[htb!]
\includegraphics[height=6cm,width=8cm]{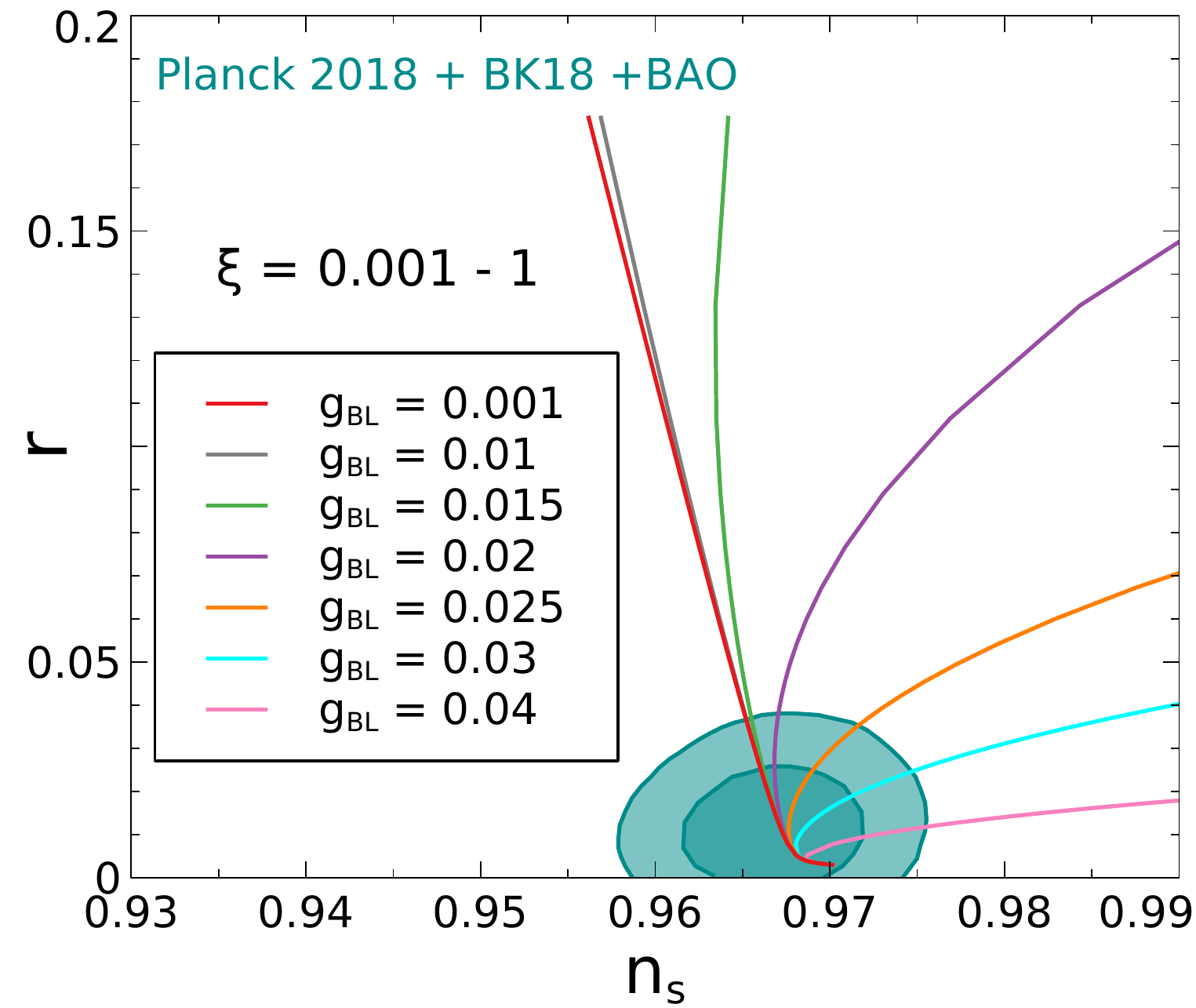}~~~~
\includegraphics[height=6cm,width=8.2cm]{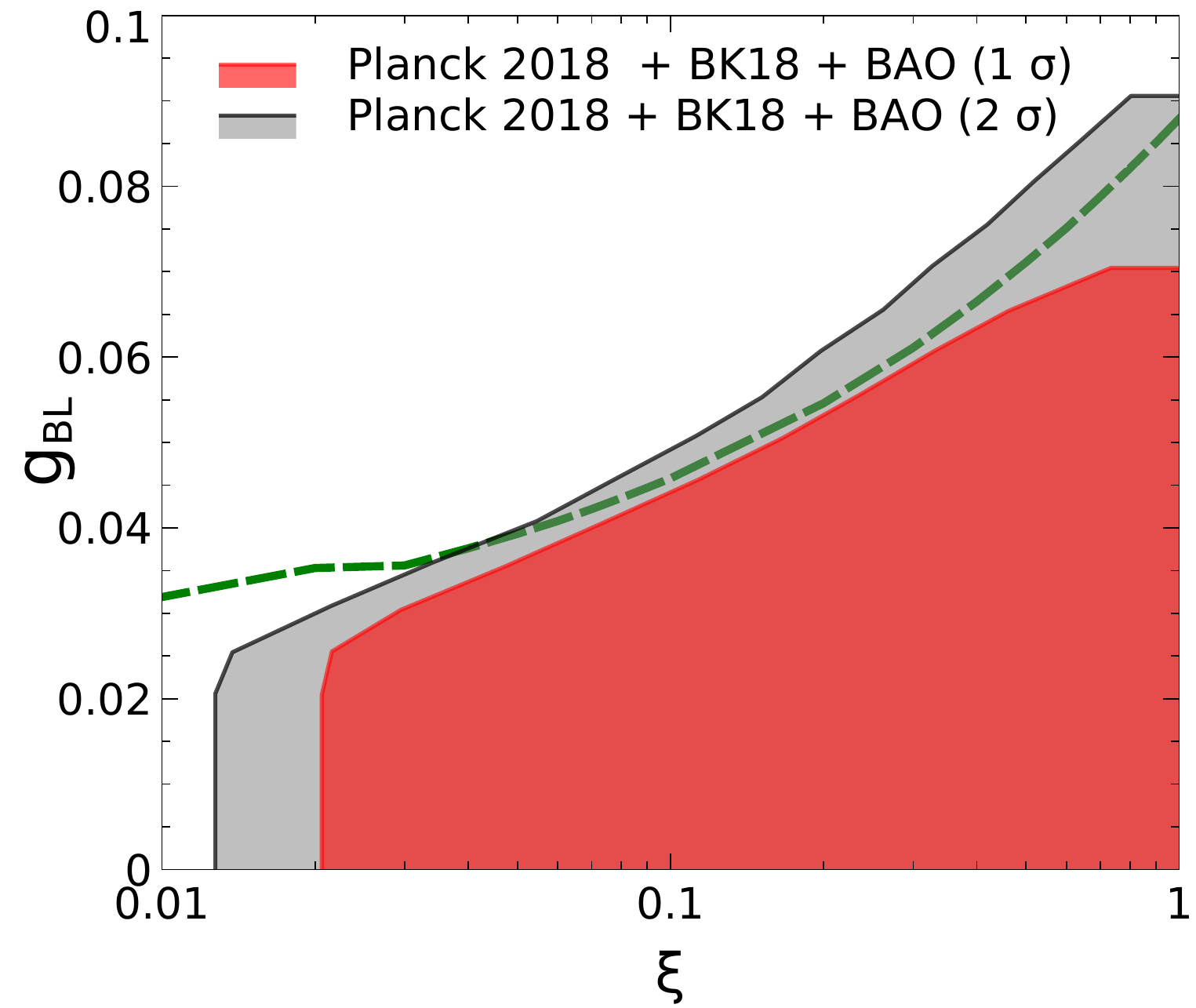}\\
\caption{[Left]: The predictions of the $B-L$ inflation in $n_s-r$ plane varying $\xi$ for different $g_{BL}$ values have been presented. The allowed $1\sigma$ and $2\sigma$ contours in the same plane from the most recent Planck 2018+BICEP/Keck analysis \cite{BICEPKeck:2021gln} are also included. [Right] We show the favored region in $\xi-g_{BL}$ plane satisfying the $1\sigma$ and $2\sigma$ bounds as provided by Planck 2018+BICEP/Keck data. The green dashed line demonstrates the upper bound on $g_{BL}$ originating from appearance of local minimum along the inflationary trajectory.}
\label{fig:ns-r}
\end{figure*}

Following the standard prescription, we use the following conformal transformation to write the action $S_J$ in the Einstein frame~\cite{Capozziello:1996xg, Kaiser:2010ps} as
\begin{equation}
\hat{g}_{\mu\nu}=\Omega^{2}g_{\mu\nu},~~\sqrt{-\hat{g}}=\Omega^{4}\sqrt{-g},
\end{equation}
so that it resembles a regular field theory action of minimal gravity. In the above equation, $\hat{g}$ represents the metric in the Einstein frame. Furthermore, to make the kinetic term of the inflaton appear canonical, we transform the $\phi$ by
\begin{equation}
\frac{d\chi}{d\phi}=\sqrt{\frac{\Omega^{2}+\frac{6\xi^{2}\phi^{2}}{M_{P}^{2}}}{\Omega^{4}}}=Z(\phi),
\end{equation}
where $\chi$ is the canonical field. Using these inputs, the inflationary potential in the Einstein frame can be written as, 
\begin{equation}
V_{E}(\phi(\chi))=\frac{V_{J}\big(\phi(\chi)\big)}{\big (\Omega\big(\phi(\chi)\big)\big )^{4}}=\frac{1}{4}\frac{\lambda_2\phi^4}{\big (1+\frac{\xi\phi^2}{M_P^2}\big )^2},
\end{equation}
where $V_J(\phi)$ is identical to $V_{\rm Inf}(\phi)$ in Eq. (\ref{eq:PotPhi}). We then make another redefinition: $\Phi={\frac{\phi}{\sqrt{1+ \frac{\xi\phi^2}{M_P^2}}}}$ and reach at a much simpler from of $V_E$ given by
\begin{align}
 V_E(\Phi)=\frac{1}{4}\lambda_2\Phi^4.\label{eq:InfE}
\end{align}
Note that for an accurate analysis, one should work with renormalisation group (RG) improved potential and in that case, $\lambda_2$ in Eq. (\ref{eq:InfE}) will be function of $\Phi$ such that,
\begin{align}
 V_E(\Phi)=\frac{1}{4}\lambda_2(\Phi)\Phi^4.\label{eq:infpotRG}
\end{align}
{One can notice that the inflaton field $\Phi$ ($=\phi/\sqrt{1+ \frac{\xi\phi^2}{M_P^2}}$) in Eq.(\ref{eq:infpotRG}) is non-canonical. Hence we redefine the slow roll parameters as functions of the non-canonical field $\Phi$ and arrive at \cite{Okada:2015lia} ,
\begin{align}
&\epsilon(\Phi)=\frac{M_{P}^{2}}{2Z(\Phi)^{2}}\Bigg(\frac{V_{E}'(\Phi)A(\Phi)}{V_{E}(\Phi)}\Bigg)^{2},\\
&\eta(\Phi)=\frac{M_{P}^{2}}{Z(\Phi)^{2}}\Bigg(\frac{V_{E}''(\Phi)A(\Phi)^2}{V_E(\Phi)}+\frac{V^\prime_E(\Phi)B(\Phi)}{V_{E}(\Phi)}\bigg),
\end{align} 
where $^\prime$ indicates the differentiation of the relevant quantity with respect to $\Phi$ and
\begin{align}
   & A(\Phi)=\left(1-\frac{\xi \Phi^2}{M_P^2}\right)^{3/2},\\
   & B(\Phi)=A(\Phi)A^\prime(\Phi)-\frac{Z'(\Phi)A(\Phi)^2}{Z(\Phi)}.
\end{align}
The number of e-folds ($N_e$) is given by 
\begin{align}
    &N_e=\int_{\Phi_{t}}^{\Phi_{\rm end}}\frac{Z(\Phi)^{2}V_{E}(\Phi)}{V_{E}'(\Phi)A(\Phi)^2}\frac{{d}\Phi}{M_{\mathrm{P}}}
\end{align} 
where $\Phi_{t}$ and $\Phi_{\rm end}$ are the inflaton field values at horizon exit and end of inflation respectively.}

{Using these expressions, we obtain the predictions for the inflationary observables namely the magnitude of spectral index ($n_s=1-6\epsilon+2\eta$) and tensor to scalar ratio ($r\sim 16\epsilon$). We also consider the number of e-folds ($N_e$) as 60.}

The full set of one loop renormalisation group evolution (RGE) equations of the relevant parameters associated with the inflationary dynamics can be found in \cite{Okada:2015lia}.
We consider a diagonal RH neutrino mass matrix with the hierarchy $M_{N_1},\,M_{N_2} \ll M_{N_3}$ \footnote{This choice is motivated from the fact that a heavier RH neutrino assists in obtaining larger amount of late entropy production and thus more effective in diluting the DM relic without violation the BBN bound.}. This implies $Y_{N_1}, Y_{N_2} \ll Y_{N_3}$ where $M_{N_i}=\frac{Y_{N_i}v_{BL}}{2}$. A simplified form for the RGE equation of $\lambda_2$ can be written by assuming $Y_{N_{3}},g_{BL}\gg \lambda_2,\lambda_3$ leading to the following beta function,
\begin{align}
 \beta_{\lambda_2}\simeq 96g_{BL}^{4}-Y_{N_3}^{4}.\label{eq:RG2}
\end{align}
Below we provide the three important conditions to realise a successful RGE improved inflation in minimal gauged $B-L$ model.

$\bullet$ In general, for $\xi\lesssim 1$, the self-quartic coupling of inflaton must be very small in order to be in agreement with the experimental bounds on inflationary observables \cite{Okada:2010jf}. Since $\lambda_2$
is very small, any deviation of $\beta_{\lambda_2}$ from zero (due to larger $g_{BL}$ and $Y_{N_3}$), if sufficient, may cause sharp changes in $\lambda_2$ value from its initial magnitude during the RGE running. This may trigger unwanted instability to the inflationary potential \cite{Okada:2015lia,Borah:2020wyc}. Therefore keeping the $\beta_{\lambda_2}$ in the vicinity of zero during inflation is a desired condition for successful inflation. To ensure $\beta_{\lambda_2}\sim0$, the equality $96g_{BL}^{4}-Y_{N_3}^4 \sim 0$ needs to be maintained. 

$\bullet$ In addition to the stability ($\beta_{\lambda_2}\sim 0$) confirmation, the inflationary potential should also be monotonically increasing function of the inflaton field value which implies $M_P\frac{d\beta_{\lambda_2}}{d\Phi}>0$ during inflation. It is reported in \cite{Okada:2015lia,Borah:2020wyc} that
with the increase of $g_{BL}$, a local minimum appears (due to the violation of $M_P\frac{d\beta_{\lambda_2}}{d\Phi}>0$) within the inflationary trajectory, which can stop the inflaton from rolling.
This poses an upper bound on the size of $B-L$ gauge coupling.

$\bullet$ Finally, another constraint on $g_{BL}$ comes from the criteria that the inflationary predictions ({\it e.g.} spectral index and tensor to scalar ratio) stay within the $1-\sigma$ allowed range as provided by Planck+BICEP experiment. This particular bound on $g_{BL}$ is stronger than the one arising due to appearance of a local minimum along the inflationary trajectory as we shall show in a while.

Next, we use the standard definitions of slow roll parameters ($\epsilon$ and $\eta$) \cite{Linde:2007fr} while calculating the magnitude of spectral index ($n_s=1-6\epsilon+2\eta$) and tensor to scalar ratio ($r\sim 16\epsilon$). We consider the number of e-folds ($N_e$) as 60. We also impose the condition $\beta_{\lambda_2}=0$ at the scale of horizon exit of inflation while estimating the inflationary observables.

We perform a numerical scan over $g_{BL}$ and $\xi$ to estimate the inflationary observables $n_s$ and $r$ considering $\beta_{\lambda_2}= 0$. We have observed that the factor $\lambda_2$ does not alter the slow roll parameters, rather it is fixed by the measured value of scalar perturbation spectrum ($P_S=2.4\times 10^{-9}$) at horizontal exit of inflaton.
It also turns out that the value of $r$ does not change much with the variation of
$g_{BL}$ for a constant value of $\xi$ since $\beta_{\lambda_2} \sim 0$ at inflationary energy scale. Contrary to this, value of $n_s$
is quite sensitive to $g_{BL}$ as it involves second order derivative of the inflationary potential. In the left panel of Fig. \ref{fig:ns-r}, we show the $n_s-r$ predictions for different $g_{BL}$ lines (with varying $\xi$) and check its viability against the improved version of Planck+BICEP/Keck data \cite{BICEPKeck:2021gln} published in 2021. In the right panel of Fig. \ref{fig:ns-r},  we constrain the $\xi-g_{BL}$ plane by using the criteria of yielding correct values for $n_s$ and $r$ allowed by the combined Planck+BICEP/Keck data \cite{BICEPKeck:2021gln}. The maximum permitted value (green dashed line) of $g_{BL}$ as function of $\xi$ is also depicted in the same figure which corresponds to non-appearance of any local minimum along the inflationary trajectory. The parameter $\lambda_2$ at inflationary energy scale can be fixed with the observed value of scalar perturbation spectrum as earlier mentioned. As an example, we find $\lambda_2^{\rm inf}\simeq \mathcal{O}(10^{-10})$ considering $\xi=1$. We shall use this particular reference point in our DM analysis.

After inflation ends, the oscillation regime of inflaton starts and the universe enters into the reheating phase. Since the minute details of reheating phase is not much relevant for the present work, we discuss it briefly here. We followed the instantaneous perturbative reheating mechanism and compute the reheating temperature. In principle, such description is not fully complete since the particle production from inflaton can also happen during very early stages of oscillation which is known as preheating mechanism. In general, this approach always makes the true reheating temperature larger than what we find by using the perturbative approximation. The estimate of reheating temperature is important to ensure that it is bigger than the mass scale $M_{Z_{BL}}$ since we talk about thermal dark matter and thermalisation of few other heavier particles (with masses $\lesssim M_{Z_{BL}}$) as well. Therefore, for any parameter space, if we can satisfy the thermalisation criteria using the perturbative approach, it automatically implies that a more involved analysis of reheating dynamics would not change our conclusion. We adopt this conservative approach in our work. The inflaton can decay to SM fields owing to its small mixing with the SM Higgs. In addition, inflaton decaying to BSM fields ($Z_{BL},N_i$) is also possible if kinematically allowed, with these BSM fields further decaying into SM fields. We have found that for the concerned ranges of the relevant parameters in DM phenomenology (to be discussed in upcoming sections), the reheating temperature always turns out to be $\mathcal{O}(10^7)$\,GeV  with $\xi=1$ and $\lambda_3\sim \mathcal{O}(10^{-7})$.

\section{WDM relic}
\label{sec:DM}
In this section, we discuss the details of WDM relic calculation. As mentioned before, keV scale dark matter $\chi$ gets thermally overproduced requiring late entropy injection from heavier RHNs (we consider late decay of $N_3$). Since entropy is not conserved, this requires us to solve coupled Boltzmann equations for dark matter candidate $\chi$, the heavier right-handed neutrino $N_3$ and temperature of the universe ($T$) which can be written as function of scale factor $a$ as follows \cite{Scherrer:1984fd,Arias:2019uol}, 
\begin{align}
& \frac{d E_{\chi}}{da}=\frac{\langle\sigma v\rangle_{\chi}}{Ha^{4}}\left((E_{\chi}^{\rm eq})^{2}-E_{\chi}^{2}\right) \, , \label{eq:bol1}\\
&\frac{d E_{N_{3}}}{da}=\frac{\langle\sigma v\rangle_{3}}{Ha^{4}}\left((E_{N_{3}}^{\rm eq})^{2}-E_{N_{3}}^{2}\right)-\frac{\Gamma_{N_3}}{Ha}E_{N_{3}} \, , \label{eq:bol2}\\
&\frac{dT}{da}=\left(1+\frac{T}{3 g_{*s}}\frac{dg_{*s}}{dT}\right)^{-1}\left[-\frac{T}{a}+\frac{\Gamma_{N_3}M_{N_3}}{3 H ~s~ a^4}E_{N_3}\right],  
\label{eq:Boltz}
\end{align}
\noindent where the entropy density is expressed by $s=\frac{2\pi^2}{45}g_{*s}T^3$, with $g_{*s}$ being the relativistic entropy degrees of freedom. The Boltzmann equations are solved by considering $N_3$ to freeze out while being non-relativistic in a way similar to WIMP type DM belonging to CDM category. This can be ensured if $N_3$ is heavy enough $M_{N_3}\sim \mathcal{O}(M_{Z_{BL}})$.
 The quantities $E_{\chi,N_3}$ in the above equations are defined as the co-moving number densities respectively, i.e., $E_{\chi,N_3}=n_{\chi,3}a^3$ where $n_{\chi,3}$ are the usual number densities. The Hubble parameter is denoted by $H$ and $\Gamma_{N_3}$ corresponds to the decay width of $N_3$.
 The interaction cross-sections for a light dark matter ($\chi$) is dominated by the $Z_{BL}$ mediated annihilations into SM fermions \cite{Biswas:2018iny}. On the other hand, the heaviest RHN ${N_3}$ can annihilate to SM particles as well as $\{Z_{BL}Z_{BL}$, $Z_{BL}H_2$, $Z_{BL}H_2$, $H_{2}H_2, N_{1,2} N_{1,2},\chi\chi\}$ final states. We incorporate all the relevant processes in thermally averaged annihilation cross sections for both $\chi$ and $N_3$ as denoted by $\langle\sigma v\rangle_{\chi}$ and $\langle\sigma v\rangle_{3}$ respectively. Here we do not include other RHNs $N_{1,2}$ considering them to decay promptly into SM leptons having negligible impact on DM phenomenology.

It should be noted that on the right hand side of Eq.(\ref{eq:bol2}) we have ignored the inverse decay term given as $\frac{\Gamma_{N_3}}{Ha}E_{N_3}^{\rm eq}$. This is due to the tiny decay width of $N_3$ which makes the inverse decay contribution to its production negligible compared to the first term on the right hand side of Eq.(\ref{eq:bol2}). The freeze-out abundance of $N_3$, therefore, is dictated primarily by the strength of $\langle\sigma v\rangle_{3}$. Also, since $N_3$ freezes out like WIMP (after becoming non-relativistic), the inverse decay contribution during the post freeze-out epoch remains highly Boltzmann suppressed. We have also verified it by explicitly including the inverse decay term in the Boltzmann equation leading to no change in our numerical results.


\begin{figure*}[htb!]
\includegraphics[height=6.5cm,width=8cm]{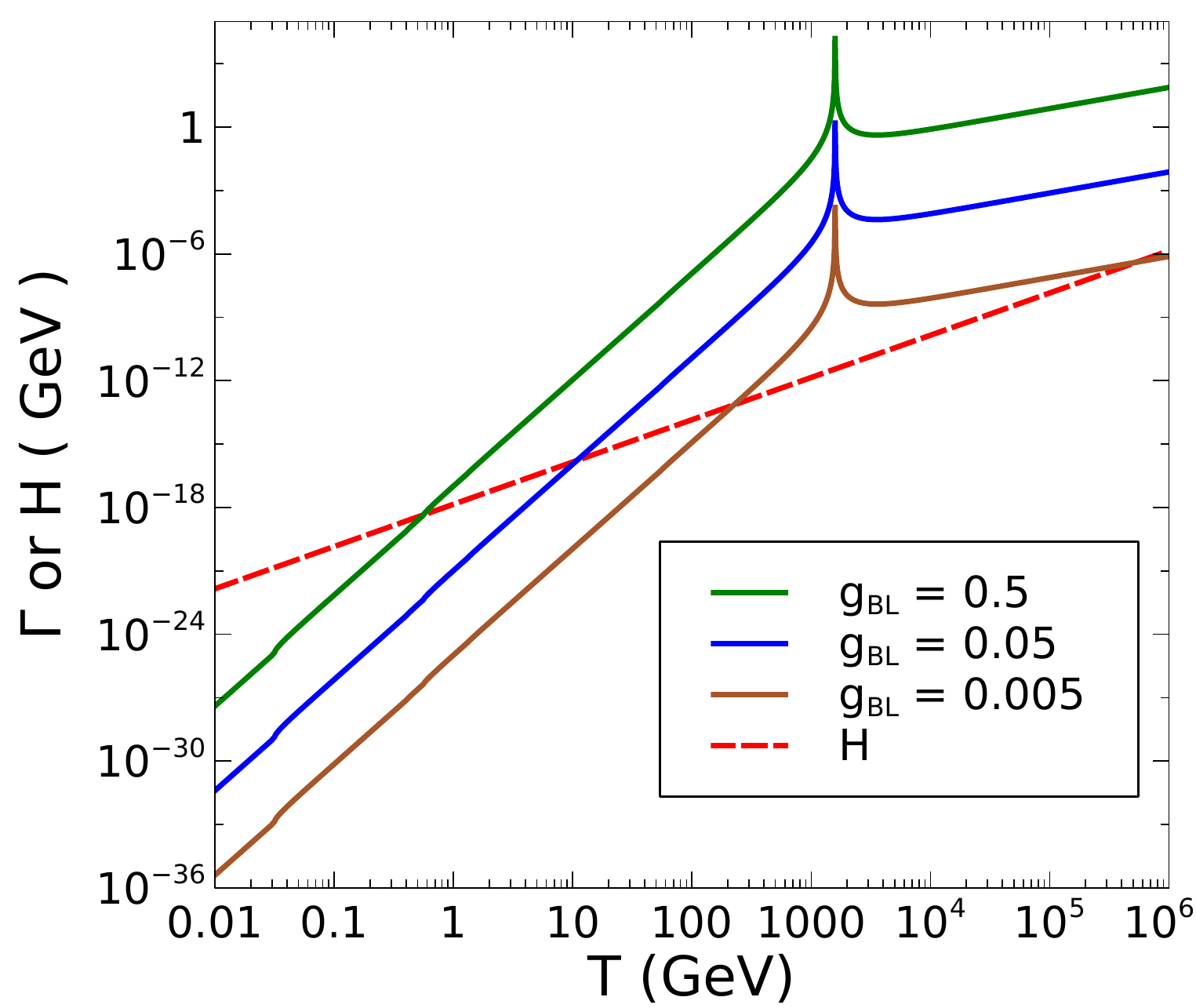}~~~~
\includegraphics[height=6.5cm,width=8cm]{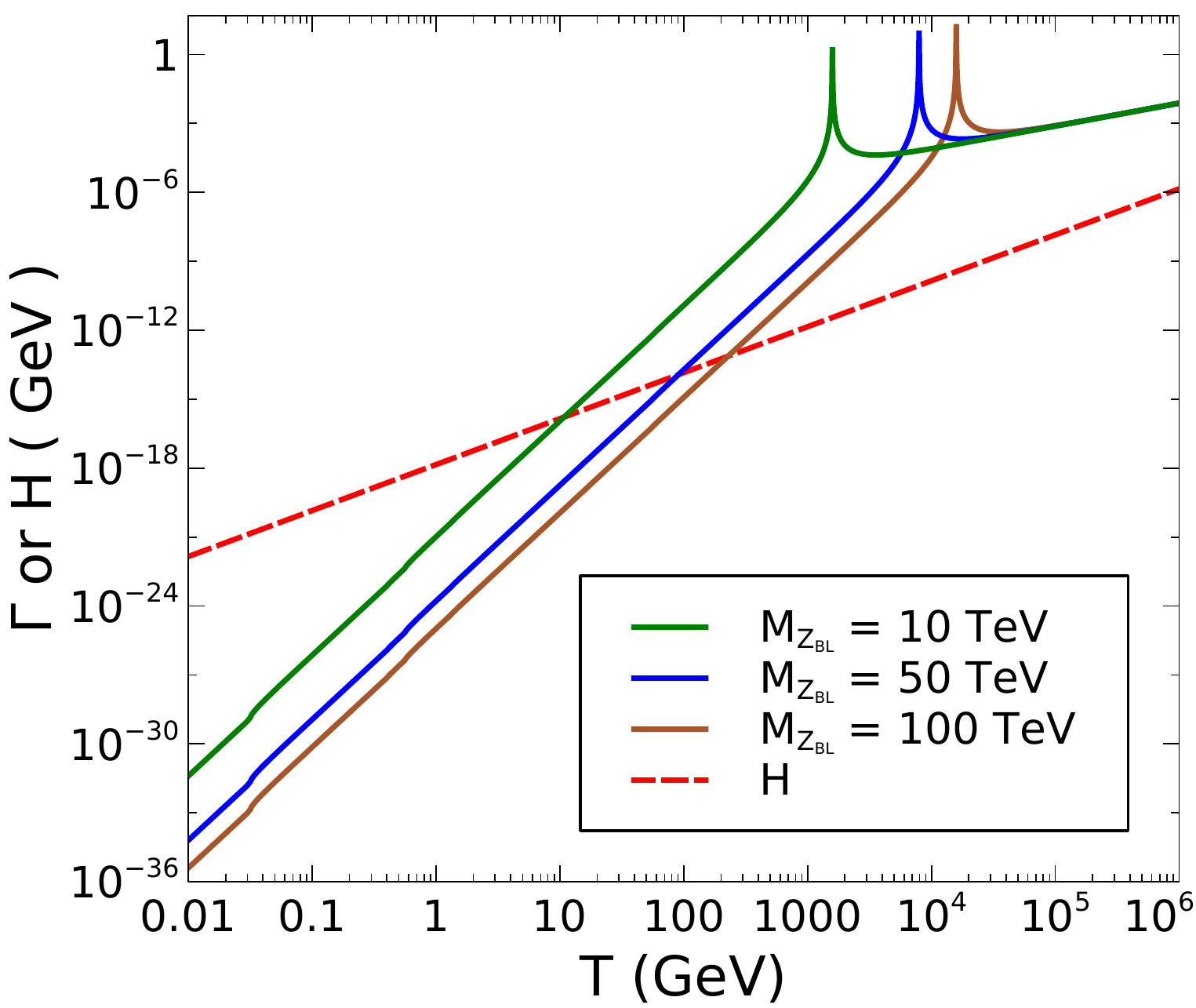}
\caption{DM interaction rate and Hubble parameter in a radiation dominated universe as function of temperature for different $g_{BL}$ values (left panel) and different $M_{Z_{BL}}$ values (right panel).}
\label{fig:GmmaH}
\end{figure*}

 The energy density and the corresponding Hubble parameter in a radiation dominated universe are defined as $$\rho_R=\frac{\pi^2}{30}g_{*\rho}T^4,~H=\sqrt{\frac{\rho_R}{3 M_P^2}}$$
 with $g_{*\rho}$ is the effective relativistic degrees of freedom and $M_P$ denotes the reduced Planck scale.
 At earlier epoch, $N_3$ was part of radiation bath by virtue of its gauge and Higgs portal interactions with the SM particles. When $N_3$ becomes non-relativistic and goes out of equilibrium it can be safely treated as matter. In that case the Hubble parameter is redefined as,
\begin{equation}
H=\sqrt{\dfrac{\frac{M_3E_{N_3}}{a^{3}}+\rho_{R}}{3M_{P}^{2}}}. \label{Hbl}
\end{equation}  
\noindent Now, the dark matter $\chi$ freezes out when it is relativistic. The decoupling temperature of $\chi$ is determined by the strength of the interactions, mainly governed by $g_{BL}$ and $M_{Z_{BL}}$. In the left panel of Fig. \ref{fig:GmmaH}, we show the evolution of the DM interaction rate ($\Gamma$) and the expansion rate of a radiation dominated universe (Hubble parameter) for different $g_{BL}$ values keeping $M_{Z_{BL}}$ fixed at 10 TeV. The right panel shows the same for varying $M_{Z_{BL}}$ values, with $g_{BL}$ being fixed at 0.05. It is clear from these plots that for lower $g_{BL}$ and higher $M_{Z_{BL}}$ values, $\chi$ decoupling occurs earlier with dark matter being relativistic, due to smaller interaction rate. Note that, lowering $g_{BL}$ also makes DM enter into equilibrium at late epochs and decreasing it beyond a particular value may not lead to its thermalisation at all.

 \begin{figure*}[htb!]
  \includegraphics[height=6.3cm,width=8cm]{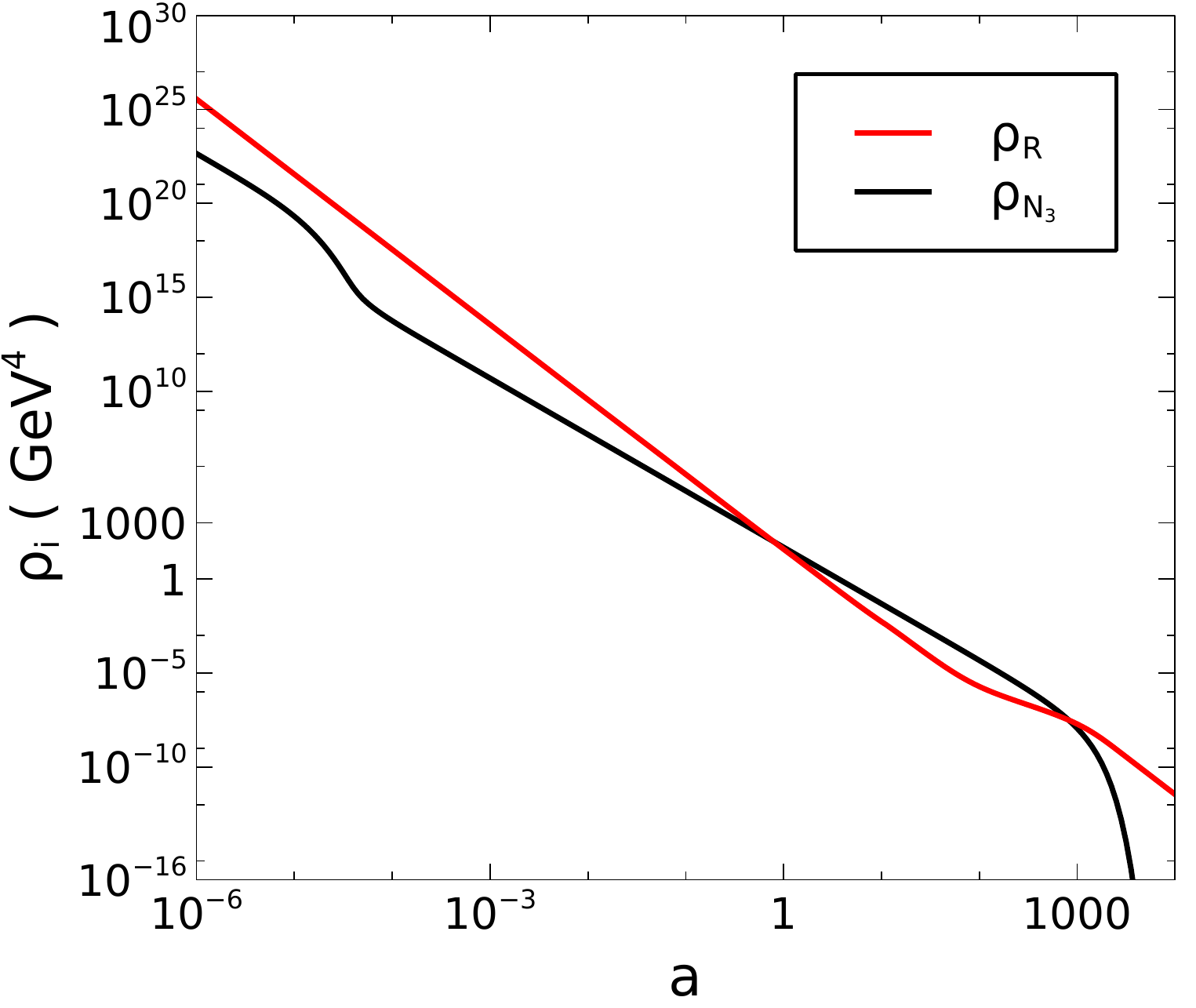}~~~~
\includegraphics[height=6.3cm,width=8cm]{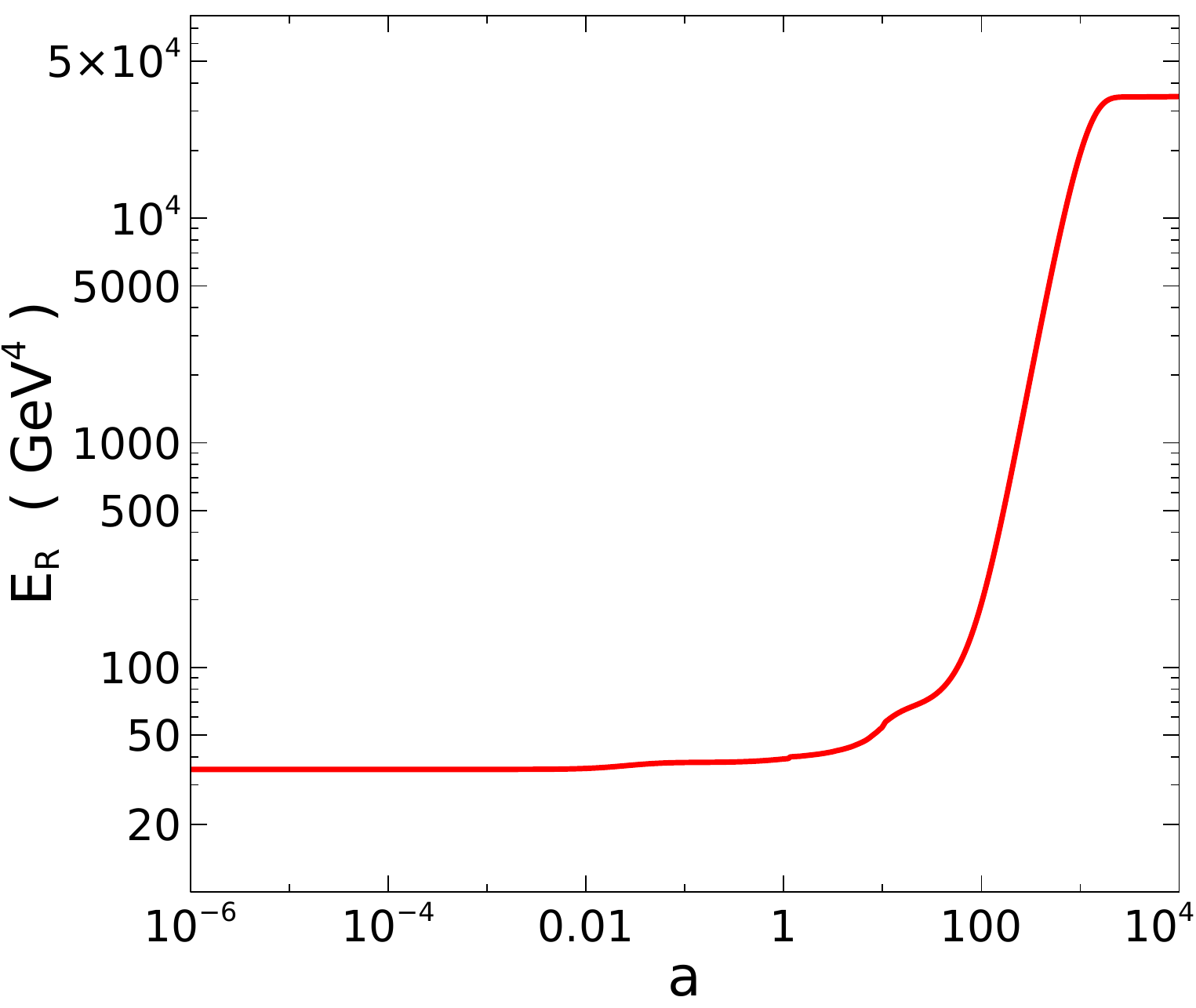}
\caption{Left panel: A comparison between the energy densities of radiation and $N_3$ are shown from early time to late epoch. Right panel: Enhancement of radiation energy density is shown due to $N_3$ decay at late time.}
\label{fig:BPplots1}
 \end{figure*}

The DM relic for the case of a relativistic freeze-out considering radiation dominated universe is simple to compute, and can be estimated as \cite{Kolb:1990vq}
\begin{equation}
\Omega_{\rm DM}h^2=76.4\times \left(\frac{3g}{4g_{*s}(x_f)}\right)\left(\frac{M_{\rm DM}}{\rm keV}\right) \, , \label{DMrlc}
\end{equation}
where $g$ is the number of internal degrees of freedom of the DM candidate, $g_{*s}(x_f)$ represents the number of relativistic entropy degrees of freedom at the instant of $\chi$ decoupling $x_f=\frac{M_{\rm DM}}{T_f}$ , where $T_f$ is the freeze-out temperature. Considering the DM to decouple around the EW scale (with $g_{*s}(x_f)\approx 106.75$), we find (for $M_{\rm DM}=5$\,keV)
\begin{equation}
\Omega_{\rm DM}h^2=76.4\left(\frac{3\times 4}{4\times 106.75}\right)\left(\frac{M_{\rm DM}}{\rm keV}\right)\approx 10 \, . \label{DMrlc2}
\end{equation}
Thus, DM remains overproduced roughly by two orders of magnitude or by a factor of around 100. This overabundance can be brought down by the late decay of $N_3$ after the DM freeze-out, which injects entropy ($s$) into the thermal bath \cite{Scherrer:1984fd}. In order to realise this possibility of sufficient entropy dilution, it is necessary for the long-lived $N_3$ to dominate the energy density of the universe at late time over the radiation component. Importantly, the validity of Eq.(\ref{DMrlc}) breaks down if the universe has such non-standard history where $N_3$ dominates the energy density of the universe for some epoch (similar to early matter domination scenarios \cite{Allahverdi:2021grt}). In that case, we need to solve the system of Boltzmann equations numerically (for analytical approach, see \cite{Bhatia:2020itt}) and obtain the relic abundance of dark matter using 
\begin{align}
 \Omega_{\rm DM} h^2=2.745\times 10^{8} M_{\rm DM}Y_{a\rightarrow \infty},
\end{align}
where $Y_{a\rightarrow \infty}=\frac{E_{\chi}}{s a^3}|_{a\rightarrow \infty}$ .

The cosmic inflation as discussed earlier has three fold impacts on the DM phenomenology. Firstly, we found that the additional gauge coupling $g_{BL}$ is restricted by the inflationary requirements. The strongest bound on $g_{BL}$ ($\lesssim 0.07$) comes from providing correct values of scalar spectral index $n_s$ and tensor-to-scalar ratio $r$ for $\xi=1$. Note that these bounds can be relaxed for very large values of $\xi \gg 1$ which we do not consider here in order to keep all dimensionless parameters within order one. Secondly, in order to ensure the stability of the inflationary potential, we require the condition $\beta_{\lambda_2}=96g_{BL}^4-Y_{N_3}^4\sim 0$ to hold true at inflationary energy scale where we consider $Y_{N_3}\gg Y_{N1},Y_{N2}$. This condition along with the allowed range of $g_{BL}$($\lesssim 0.07$), controls the fate of both $\chi$ and $N_3$ thermalisation and also determine the decoupling temperature or (non-relativistic) freeze-out abundance of $N_3$ which is a crucial parameter to estimate the entropy production at late epoch. 
Lastly, the $\beta_{\lambda_2}=0$ condition also connects the mass scales $M_{Z_{BL}}$ and $M_{N_i}$ by the ratio $\frac{M_{Z_{BL}}}{M_{N_3}}=\frac{\sqrt{2}g_{BL}}{Y_{N_3}}\simeq \sqrt{2}\left(\frac{1}{96}\right)^{1/4}$.

We then consider the RGE running of all the relevant couplings from inflationary energy scale upto the $B-L$ breaking scale and obtain the values for 
 $M_{N_2}$, $M_{Z_{BL}}$ and other mass parameters. It turns out that the RGE running effects on $g_{BL}$ and $Y_{N_i}$ in the range of our interest are very minimal in view of bringing any noticeable change to the DM phenomenology.

\begin{figure*}[htb!]
\includegraphics[height=6.3cm,width=8cm]{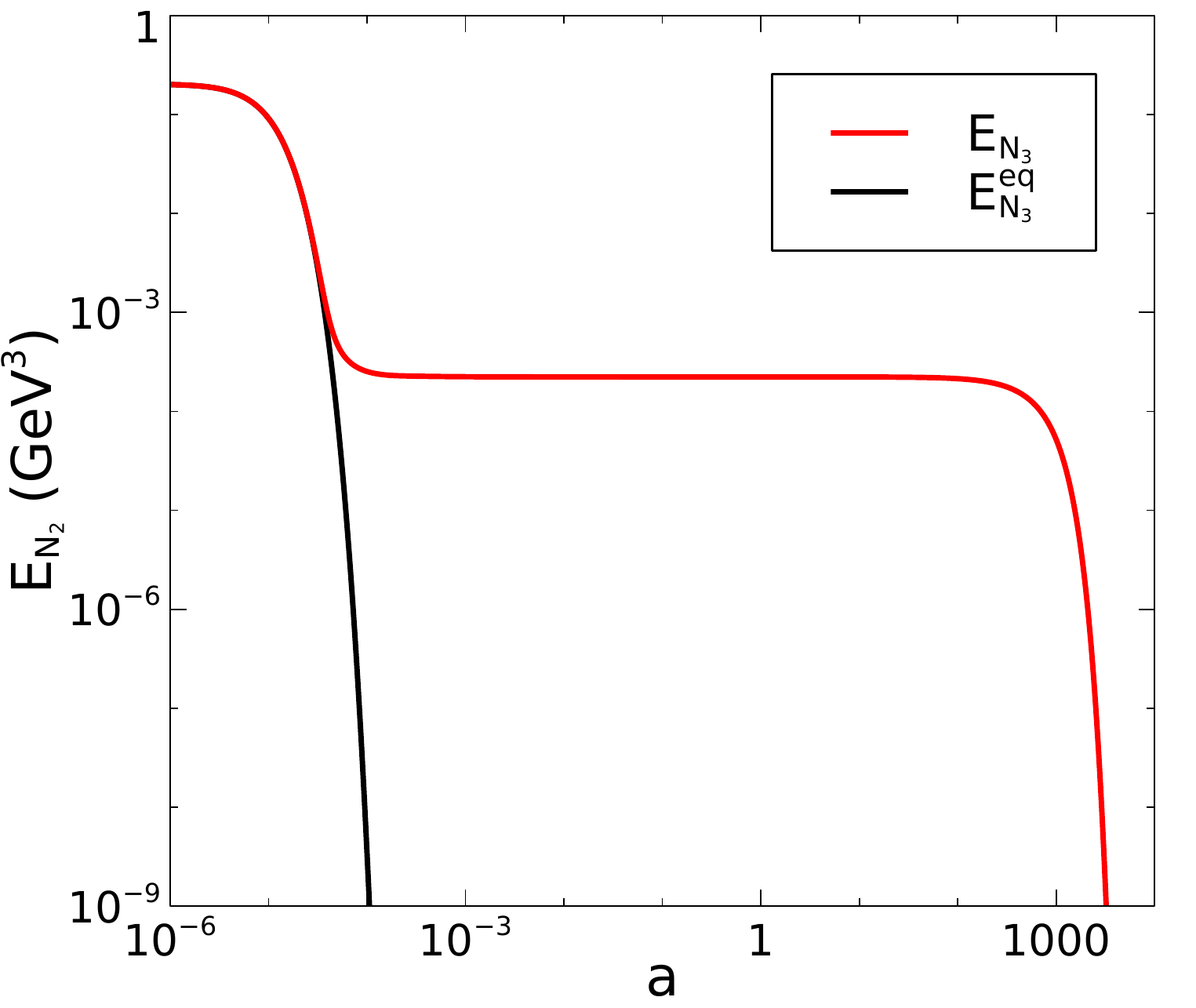}~~~~
\includegraphics[height=6.3cm,width=8.3cm]{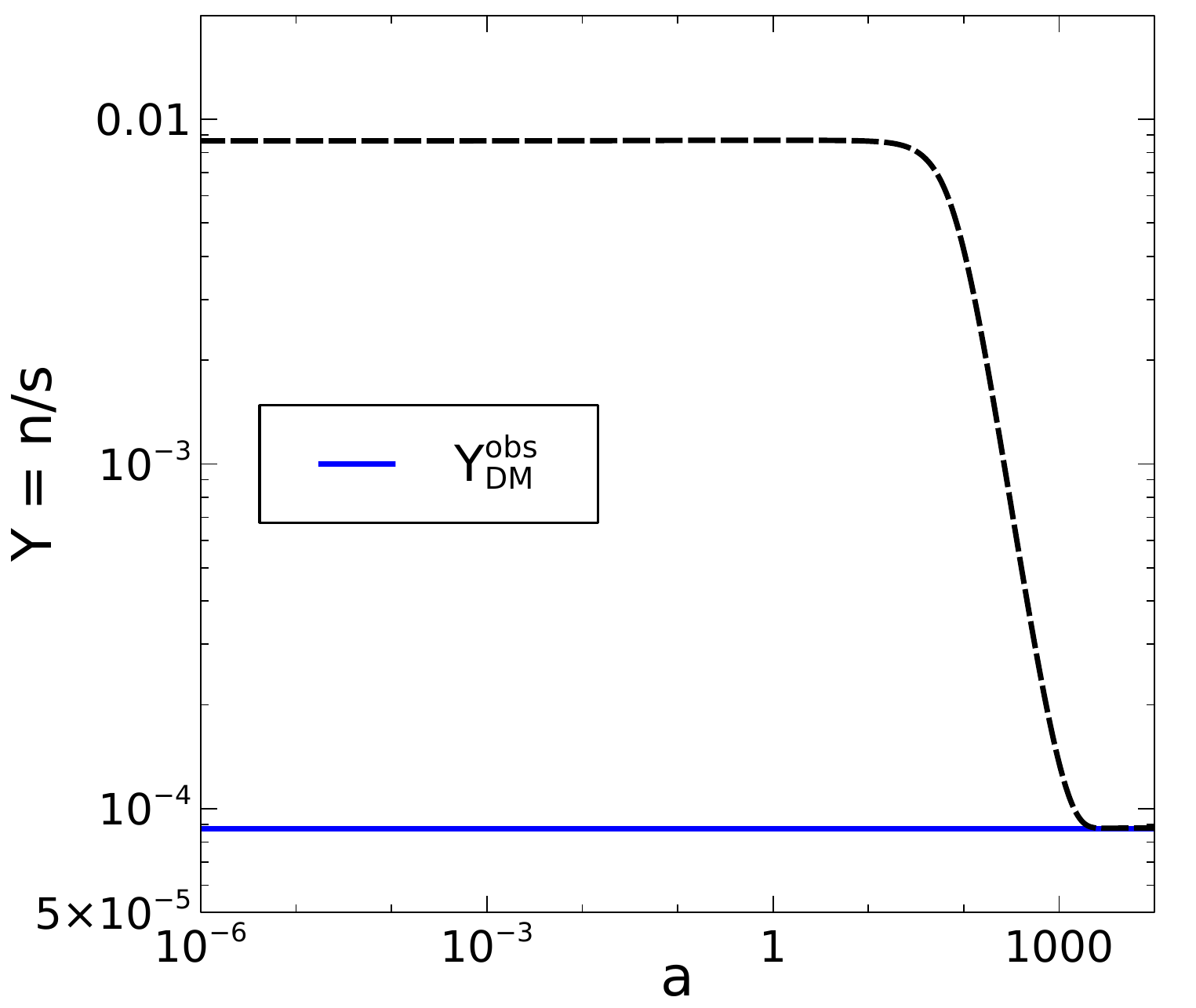}\\
\caption{Left panel: Evolution of $E_{N_3}^{\rm eq}$ (solid black) and $E_{N_3}$ (solid red) as function of scale factor. Right panel: The evolution of comoving DM density (dashed line) as function of scale factor where the effect of late time entropy dilution can be clearly observed. The blue solid line corresponds to the required comoving DM number density at present epoch from observations.}
\label{fig:BPplots2}
\end{figure*}
 
 We solve the system of Boltzmann equations (Eq. \eqref{eq:bol1}, \eqref{eq:bol2}, \eqref{eq:Boltz}) in the post-reheating era as function of the scale factor ranging from of $a_{i}$ to $a_f$, where we define $T_{i}a_{i}=1$. The relevant independent parameters which are crucial for DM phenomenology are given by,
 \begin{align}
  \Big\{g_{BL},M_{Z_{BL}}, M_{N_3},\Gamma_{N_3}\Big\}
 \end{align}
 The mass scale $M_{N_3}$ and $M_{Z_{BL}}$ are connected through the stability condition $\beta_{\lambda_2}\sim 0$ at inflationary energy scale. The scalar sector couplings are considered to be small and have little impact on the DM phenomenology except the $\lambda_2$ which fixes $M_{H_2}$. The decay of $N_3$ can occur dominantly at tree level to SM Higgs and leptons in the final states. The decay width of the $N_3$ is proportional to $|Y_D^{\alpha 3}|^2 M_{N_3}$ where we have always considered $M_{N_3}>M_{H_1}\simeq 125$ GeV.  
 
\begin{figure*}[htb!]
\includegraphics[height=6.3cm,width=8cm]{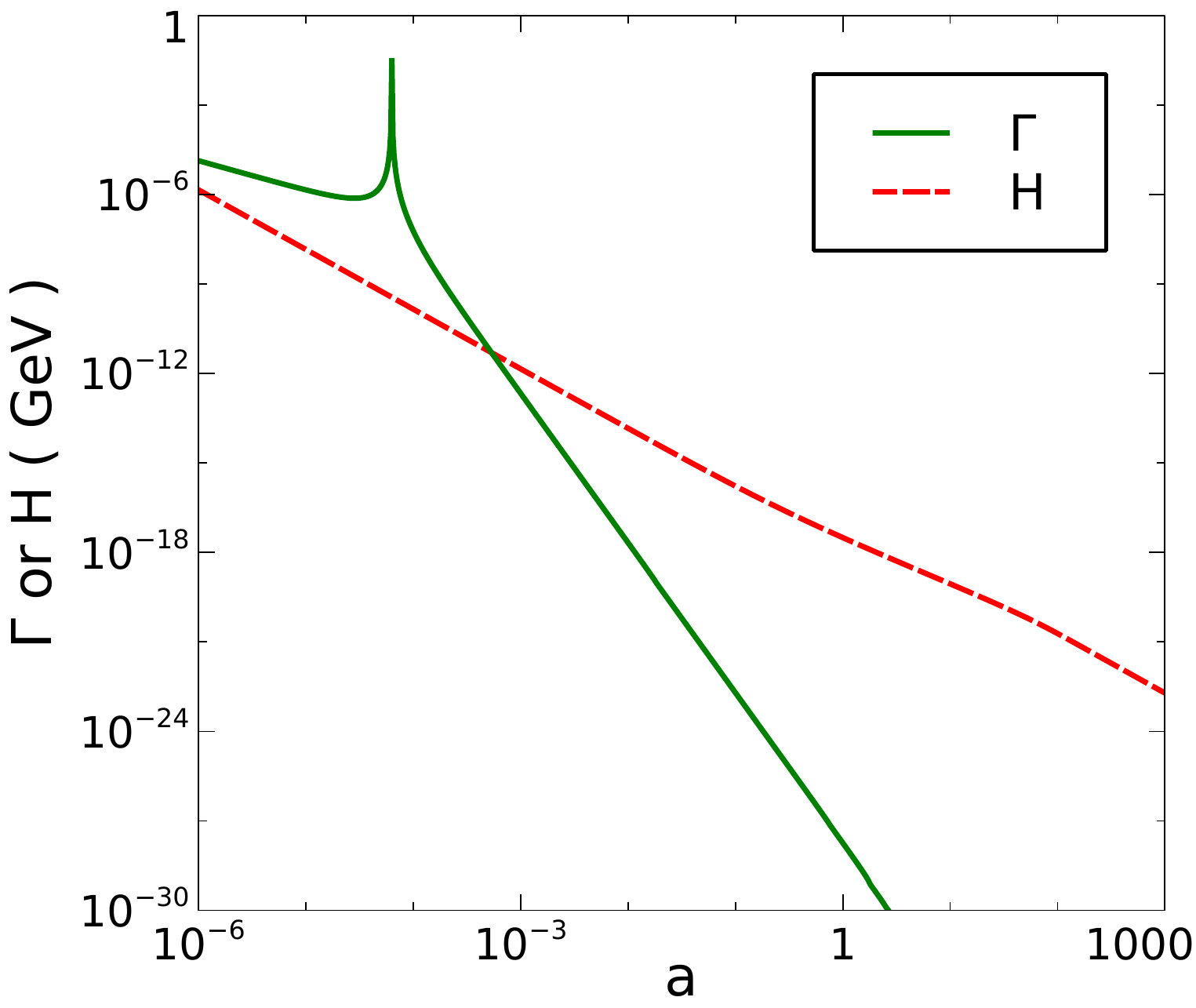}~~~~
\includegraphics[height=6.3cm,width=7.2cm]{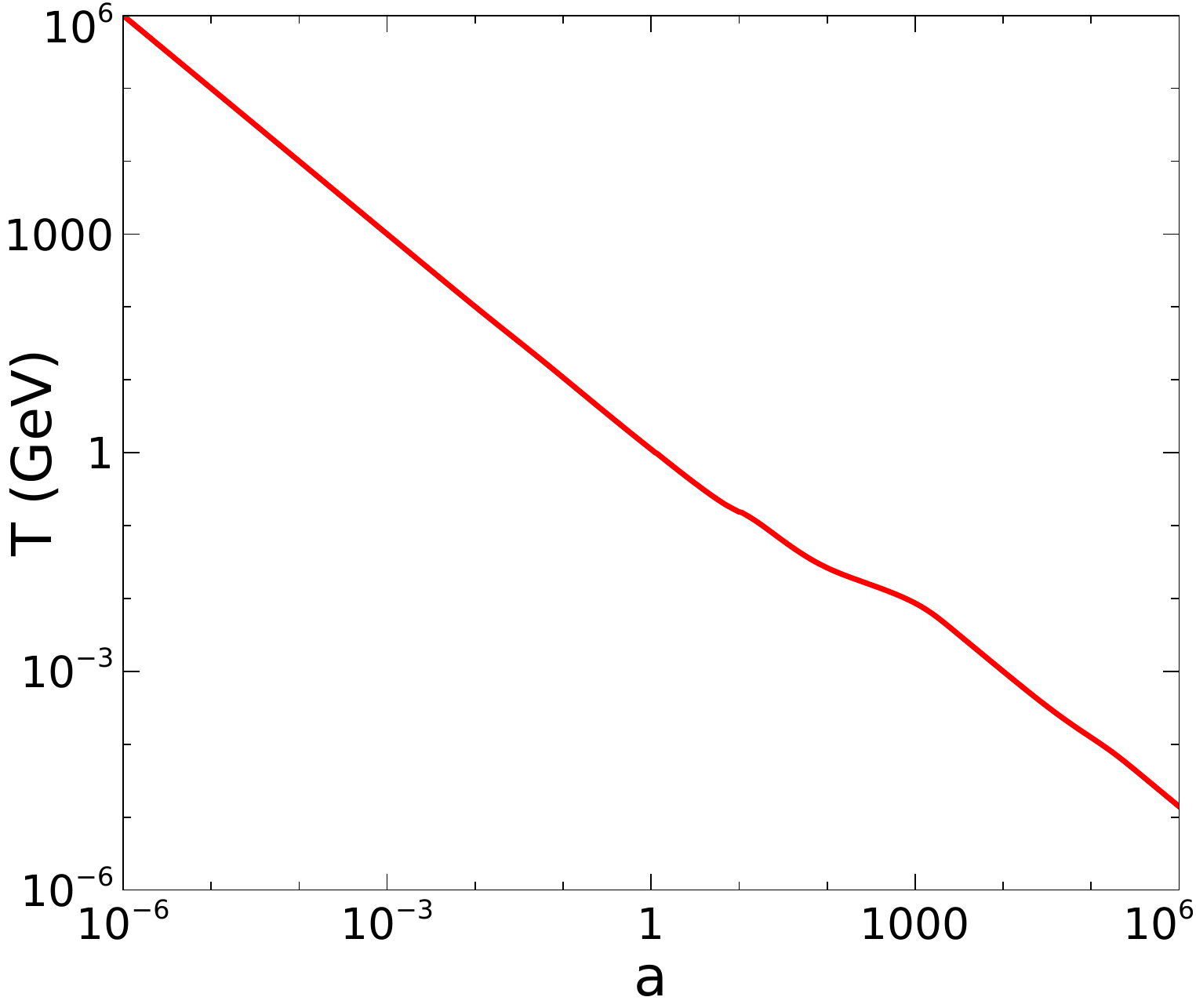}
\caption{Left panel: The interaction rate of DM $\chi$ in comparison to Hubble expansion rate of the universe with an intermediate $N_3$ dominated phase. It is seen that the DM decouples before $N_3$ dominates (see Fig. \ref{fig:BPplots1}) the energy density of the universe for the benchmark point as listed in table \ref{tab:BP1}. Right panel: The temperature evolution of the universe with an intermediate $N_3$ dominated phase.}
\label{fig:BPplots3} 
\end{figure*}   
  \begin{figure*}[htb!]
~~~~~~\includegraphics[height=7cm,width=8.6cm]{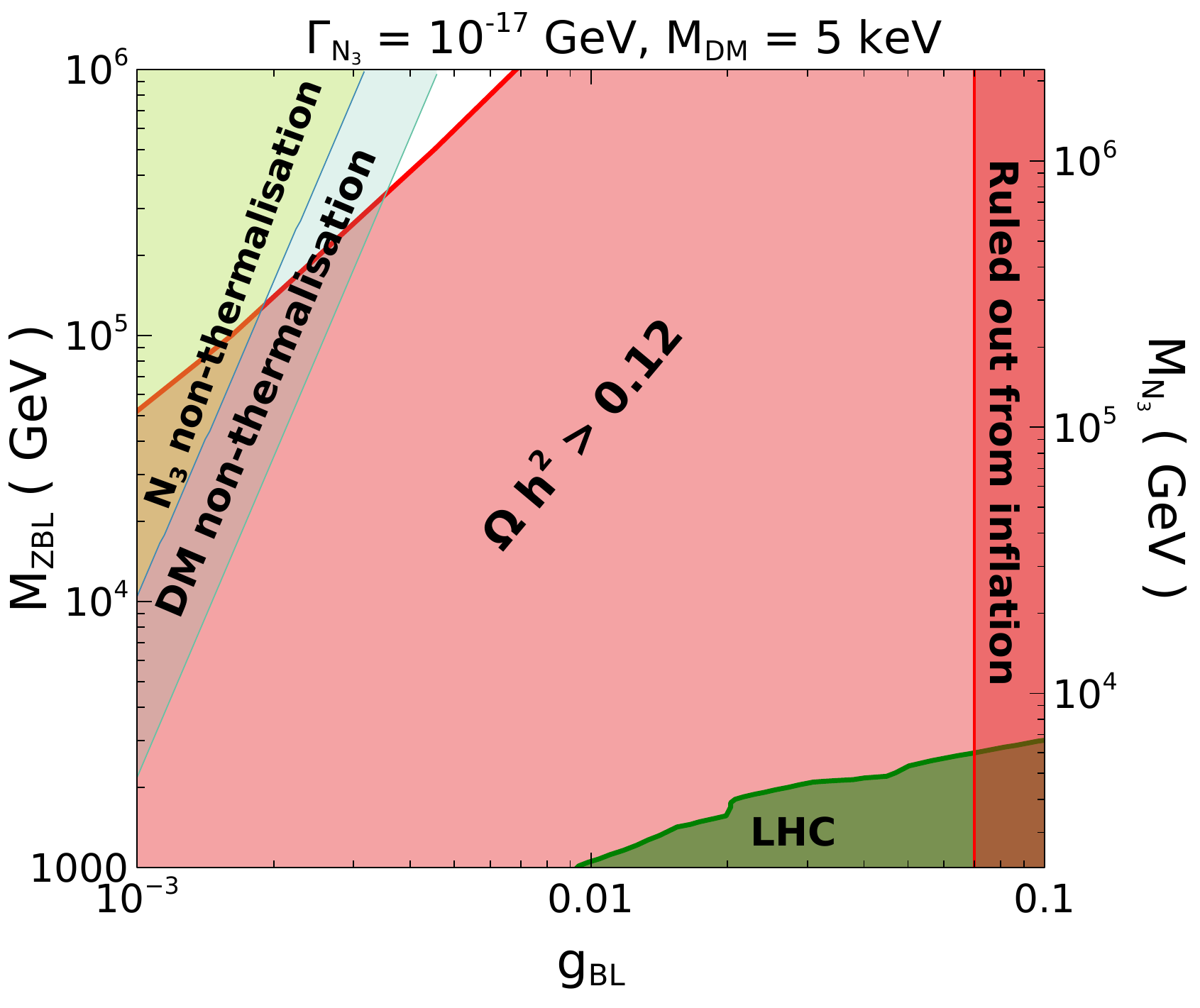}~~~~
\includegraphics[height=7cm,width=8.6cm]{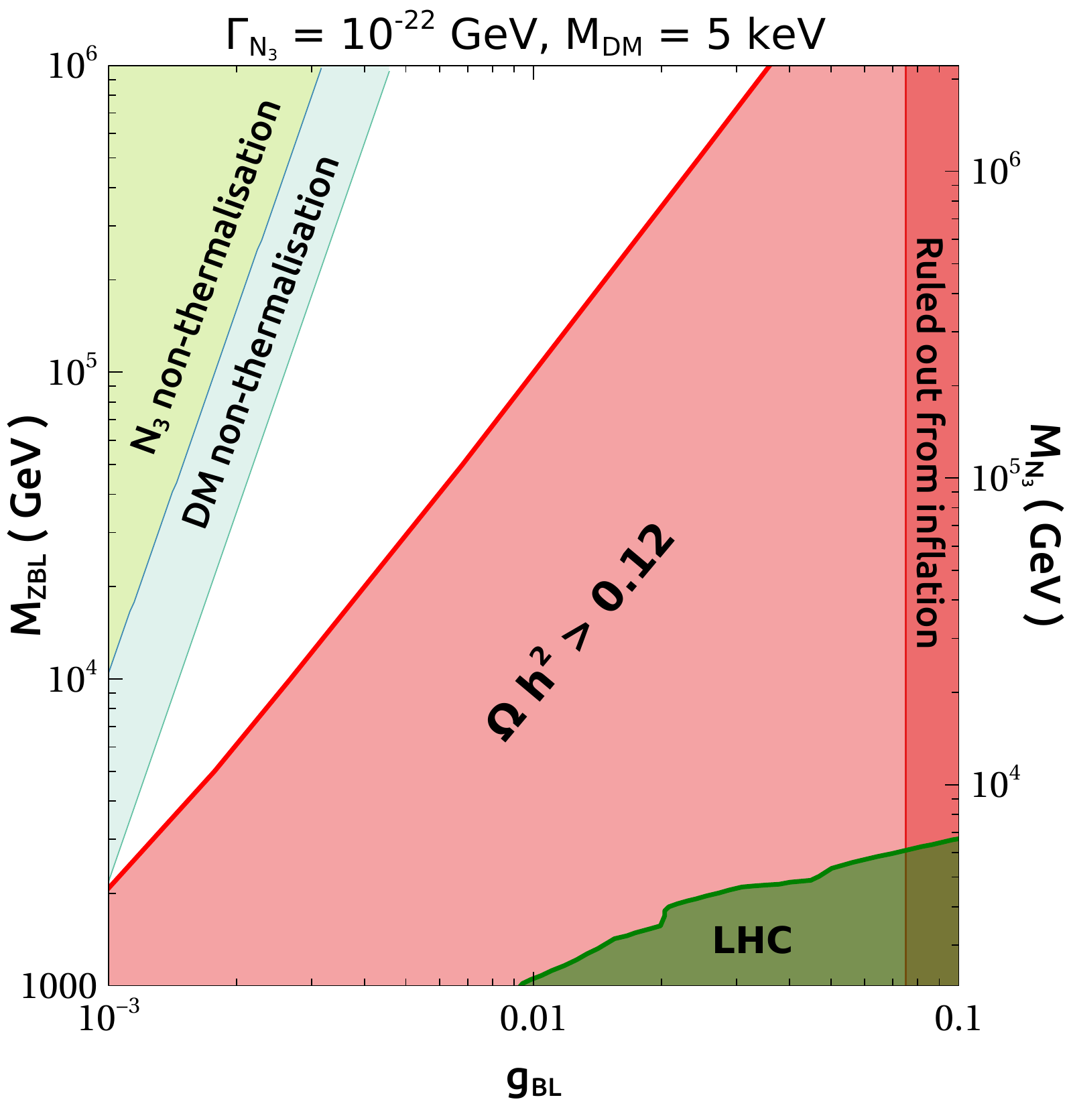}
\caption{Parameter space in $g_{BL}-M_{Z_{BL}}$ plane after imposing all relevant constraints. Only the white region is allowed leading to observed DM relic $\Omega_{\chi} h^2\lesssim 0.12$. Decay width of $N_3$ is fixed at $\Gamma_{N_3}=10^{-17}$ GeV (left panel), $\Gamma_{N_3}=10^{-22}$ GeV (right panel).}
\label{fig:MZp-gBL}
\end{figure*}    

We first stick to a benchmark scenario (as given in table \ref{tab:BP1}) to explain the dynamics of the DM phenomenology.   
\begin{table}
	\begin{center} 
			\begin{tabular}{cc}
				\hline
				\hline 
				~~~Parameters~~~  & ~~~~~~~~~Values~~~~~~~~~  \\
				\hline
				$g_{BL}$ &  0.01 \\ 
    $M_{Z_{BL}}$  &  100 TeV \\ 
    $M_{DM}$ & 5 keV \\
    $\Gamma_{N_3}$ &  $10^{-22}$ GeV\\ 
				\hline
				\hline 
			\end{tabular}
	\end{center}
	\caption{Benchmark point used for Figs.\,\ref{fig:BPplots1}}- \ref{fig:BPplots3}. 
		\label{tab:BP1}
\end{table} 
Recall that the Yukawa coupling $Y_{N_3}$ (or mass $M_{N_3})$ can not be chosen arbitrarily since it is fixed by $\beta_{\lambda_2}= 0$ condition at inflationary energy scale. The solution of the coupled Boltzmann equations for the benchmark point in table \ref{tab:BP1} are shown in Figs. \ref{fig:BPplots1}-\ref{fig:BPplots3}.

$\bullet$ The left panel of Fig.\,\ref{fig:BPplots1} shows the variation of radiation and $N_3$ energy densities from early to late epochs. Initially the radiation dominates over $N_3$ abundance as usual, followed by an intermediate epoch where $N_3$ starts dominating. The radiation domination is recovered again after $N_3$ starts decaying into SM particles. The enhancement of radiation energy density ($E_R=\rho_R a^4$) from out-of-equilibrium decay of $N_3$ at late epoch can also be observed from the right panel of Fig. \ref{fig:BPplots1}.

$\bullet $ In the left panel of Fig. \ref{fig:BPplots2}, evolution of $E_{N_3}$ as function of scale factor is shown. The solid black line indicates the scale factor dependence of $E_{N_3}^{\rm eq}$ while the actual abundance $E_{N_3}$ is shown by the solid red line. Initially $N_3$ was part of radiation bath and freezes out when the interaction rates goes below the Hubble rate. At late epochs (for large $a$) $N_3$ dominates the energy budget of the universe and finally decays into radiation. In the right panel of Fig. \ref{fig:BPplots2}, the evolution of comoving DM density (dashed line) is shown as function of the scale factor which goes through late time dilution due to sizeable entropy production from $N_3$ decay. For the chosen benchmark point, the final relic abundance exactly matches with the measured one by Planck, shown by the solid blue line.

 $\bullet$ In the left panel of Fig.\,\ref{fig:BPplots3}, we draw a comparison between the interaction rate of $\chi$ and the Hubble parameter by considering an intermediate $N_3$ dominated era. This clearly shows that DM reaches thermal equilibrium at early epochs, prior to $N_3$ domination. Also, the decoupling temperature of $N_3$ is larger than that of $\chi$, as can be seen by comparing with Fig. \ref{fig:BPplots2}. In addition to this, $\chi$ decouples before the $N_3$ domination in the Hubble parameter sets in. The right panel of Fig. \ref{fig:BPplots3} shows the temperature evolution with the scale factor. The temperature varies as $\left(\frac{1}{a}\right)$ before the start of $N_3$ domination as well as after the completion of $N_3$ decay with a small kink when $N_3$ decays indicating the entropy injection. In between for a brief period, different pattern of the temperature evolution is observed which is due to $N_3$ domination and the entropy injection into the radiation bath.

  \begin{figure*}[htb!]
\includegraphics[height=7cm,width=8cm]{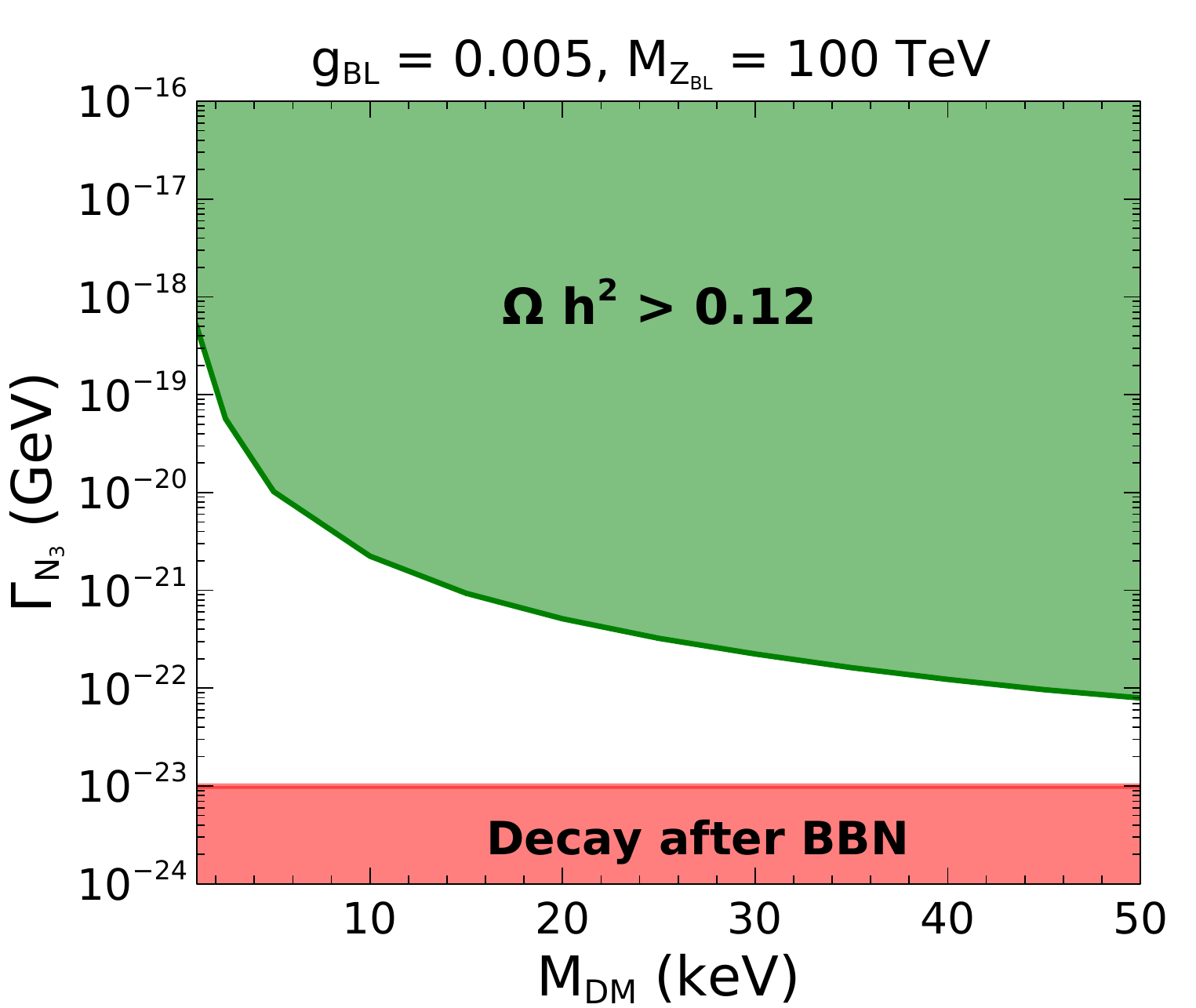}~~~~
\includegraphics[height=7cm,width=8cm]{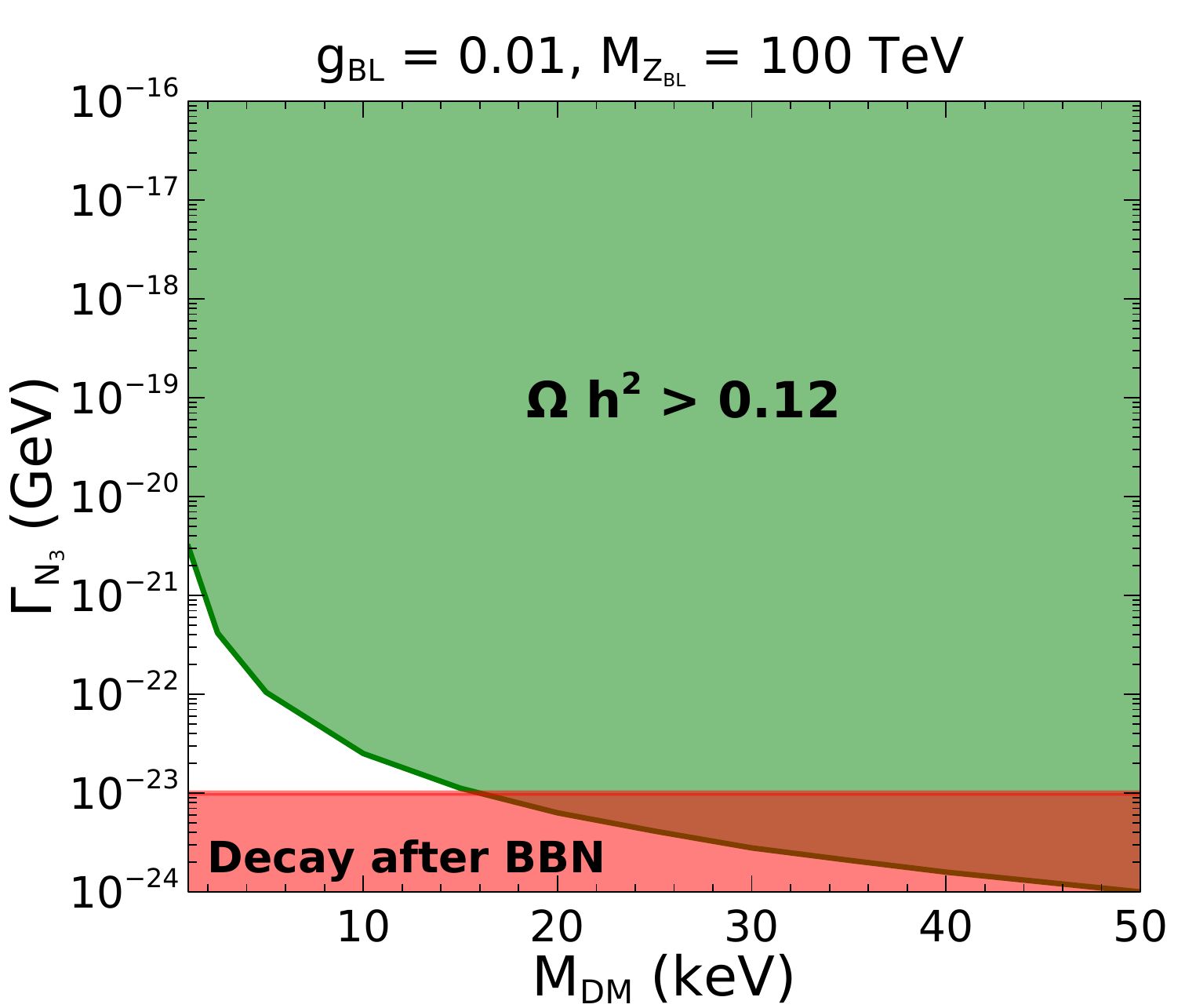}
\caption{Parameter space in $M_{\rm DM}-\Gamma_{N_3}$ plane for two sets of $(g_{BL},M_{Z_{BL}})$ values. A lower bound on $\Gamma_{N_3}$ is also shown such that $N_3$ decays completely before the BBN.}
\label{fig:Gamma-MDM}
\end{figure*}    

 We then perform a numerical scan by fixing $\Gamma_{N_3}$ at two different benchmark values to constrain the $g_{BL}-M_{Z_{BL}}$ plane from the condition of satisfying the five requirements namely, (i) $N_3$ thermalisation at early epochs, (ii) $\chi$ thermalisation at early epochs, (iii) bound on $g_{BL}$ from inflationary dynamics, (iv) LHC bounds and (v) correct final relic of DM. The resulting parameter space after applying all relevant bounds is shown in Fig. \ref{fig:MZp-gBL}. The left panel is for $\Gamma_{N_3}=10^{-17}$ GeV while in the right panel we have $\Gamma_{N_3}=10^{-22}$ GeV. The mass of $N_3$ is varied in the range of $10^3-10^6$ GeV as shown in the figure. The light blue and light green regions in the upper left corners are ruled out due to non-thermalisation of $\chi$ and $N_3$ respectively. The region corresponding to larger values of $g_{BL}$ are disfavoured by inflationary criteria mentioned earlier. The pink shaded region corresponds to DM overabundance or equivalently, insufficient entropy dilution from $N_3$ decay. The disfavored region from the LHC in the same plane is also highlighted by the brown shaded region. The relatively weaker bound from the LHC results from the search for high mass dilepton resonances. Note that the expression for $\Gamma_{N_3}$ contains $M_{N_3}$ which varies for each point (satisfying the condition $\beta_{\lambda_2}=0$) in the $g_{BL}-M_{Z_{BL}}$ plane, however a constant $\Gamma_{N_3}$ can always be obtained by tuning the neutrino Yukawa coupling $Y_D^{\alpha 3}$. We identify the allowed region (white) where the first four constraints mentioned above are satisfied and also DM relic corresponds $\Omega h^2\lesssim 0.12$. Along the boundary line between pink and white coloured regions corresponding to correct DM relic $\Omega_{\chi} h^2\sim 0.12$, larger $g_{BL}$ requires heavier $Z_{BL}$. This attributes to the fact that, for a fixed $M_{Z_{BL}}$ a larger value of $g_{BL}$ makes the $N_3$ to decouple later from the radiation bath. This leads to suppressed freeze out abundance of $N_3$ and eventually reduced amount of entropy production from its decay. To obviate that, one needs to raise $M_{Z_{BL}}$ accordingly in order to decrease the interaction cross section for $N_3$ and successively obtaining the correct relic abundance. In the same context, a larger $\Gamma_{N_3}$ (early decay of $N_3$) for constant $g_{BL}$ and $M_{Z_{BL}}$ means lower impact on dilution of DM abundance. Hence we see less amount of allowed region (white) in the left panel of Fig. \ref{fig:MZp-gBL} compared to the right panel with lower $\Gamma_{N_3}$. 
 
 To show the dependence on $N_3$ decay width further, in Fig. \ref{fig:Gamma-MDM}, we obtain the relic satisfied region ($\Omega h^2\lesssim 0.12$) in $M_{\rm DM}-\Gamma_{N_3}$ plane considering  two different sets of $(g_{BL},M_{Z_{BL}})$. The allowed space shrinks significantly for larger $g_{BL}$ since the freeze-out of $N_3$ is delayed compared to the scenario with smaller $g_{BL}$ value leading to less entropy dilution. Also, for larger DM mass, the required order of $\Gamma_{N_3}$ is smaller        
 to bring the DM abundance within the desired limit. This is due to the fact that larger DM mass yields enhanced relic and hence needs larger amount of entropy injection to the SM bath at late epochs for sufficient dilution of overproduced thermal DM relic. It is pertinent to comment here that the decay width of $N_3$ cannot be arbitrary small. Decay of $N_3$ around or after the epoch of big bang nucleosynthesis (BBN) may raise the neutrino temperature \cite{Escudero:2018mvt} which is strictly restricted by the number of relativistic degrees of freedom during BBN as measured by Planck \cite{Planck:2018vyg}. In view of this, it is safe to keep lifetime ($\tau_{N_3}$) of $N_3$ typically below one second. We have highlighted the corresponding bound on $\Gamma_{N_3}$ (red) in Fig. \ref{fig:Gamma-MDM}.     
 \vspace{2mm}
 \section{Neutrino mass and leptogenesis}\label{sec:NeuMass}
 It is clear from the above discussion that one requires very small decay width of the $N_3$ for adequate entropy production which dilutes the thermally overproduced DM relic. This, in turn, will require tiny Dirac Yukawa coupling of $N_3$ ($Y_D^{\alpha 3}$) with SM leptons. We anticipate that such smallness of $Y_D^{\alpha 3}$ can be associated with unique prediction of lightest active neutrino mass. The other entries of Dirac Yukawa matrix can satisfy the neutrino oscillation data since dynamics or interaction pattern of $N_1$ and $N_2$ are not relevant for DM phenomenology within our setup.
 
 We can write the Dirac Yukawa couplings in terms of neutrino parameters using the Casas Ibarra parametrisation \cite{Casas:2001sr}
\begin{align}
Y_D^T = \frac{\sqrt{2}}{v}\sqrt{M_N}~\mathbb{R}~\sqrt{m_{\nu}^d}~\mathcal{U}^{\dagger}\,,
\label{eq:CI}
\end{align}
\noindent where $\mathbb{R}$ is a complex orthogonal matrix with $\mathbb{R}^T \mathbb{R} = I$ and $\mathcal{U}$ represents the standard Pontecorvo–Maki–Nakagawa–Sakata (PMNS) leptonic mixing matrix. We make the following choice of $\mathbb{R}$ \cite{Ibarra:2003up}
\begin{align}
\mathbb{R} =
\begin{pmatrix}
0 & \cos{\gamma} & \sin{\gamma}\\
0 & -\sin{\gamma} & \cos{\gamma}\\
1 & 0 & 0
\end{pmatrix}\,,
\label{eq:rot-mat}
\end{align} 
in order to be consistent with feebly coupled $N_3$. In the DM phenomenology, we have worked in the limit: $Y_{D}^{\alpha 1},Y_{D}^{\alpha 2}\ll Y_{D}^{\alpha 3}$. Note that $\Gamma_{N_3}\simeq\frac{1}{16\pi}\sum_{\alpha}\big|Y_{D}^{\alpha 3}\big|^2$ \footnote{$N_3$ has two decay modes at tree level: (i) $N_3\to h \nu_\alpha$ with $\Gamma_{N_3\to h \nu_\alpha}\propto |Y_{D}^{\alpha 3}|^2 M_{N_3}$ and (ii) $N_3\to Z_{BL} \nu_\alpha$ with $\Gamma_{N_3\to Z_{BL} \nu_\alpha}\propto \frac{|Y_{D}^{\alpha 3}|^2 v^2}{M_{N_3}}g_{BL}^2$. When $M_{N_3}> v=246$ GeV and $g_{BL}\ll 1$ the
later decay process of $N_3$ is always subdominant.}. In Eq.(\ref{eq:CI}), we express the diagonal SM neutrino mass matrix $m_{\nu}^d$ as \{$m_{{\nu}_{ l}},\sqrt{(m_{{\nu}_{\rm l}}^2+\Delta m_{21}^2},\sqrt{(m_{{\nu}_{\rm l}}^2+\Delta m_{31}^2}$\} considering normal hierarchy and the lightest active neutrino mass $m_{\nu_{l}}$ as a free parameter. We have also used the best fit values for $\Delta m_{21}^2$ and $\Delta m_{21}^2$ and neutrino mixing angles \cite{ParticleDataGroup:2020ssz} with Dirac CP violating parameter $\delta_{CP}=\frac{5\pi}{2}$.  Note that the eigenvalues of the neutrino mass matrix is independent of elements of $\mathbb{R}$ matrix.

\begin{figure}[htb!]
\includegraphics[height=6.6cm,width=9cm]{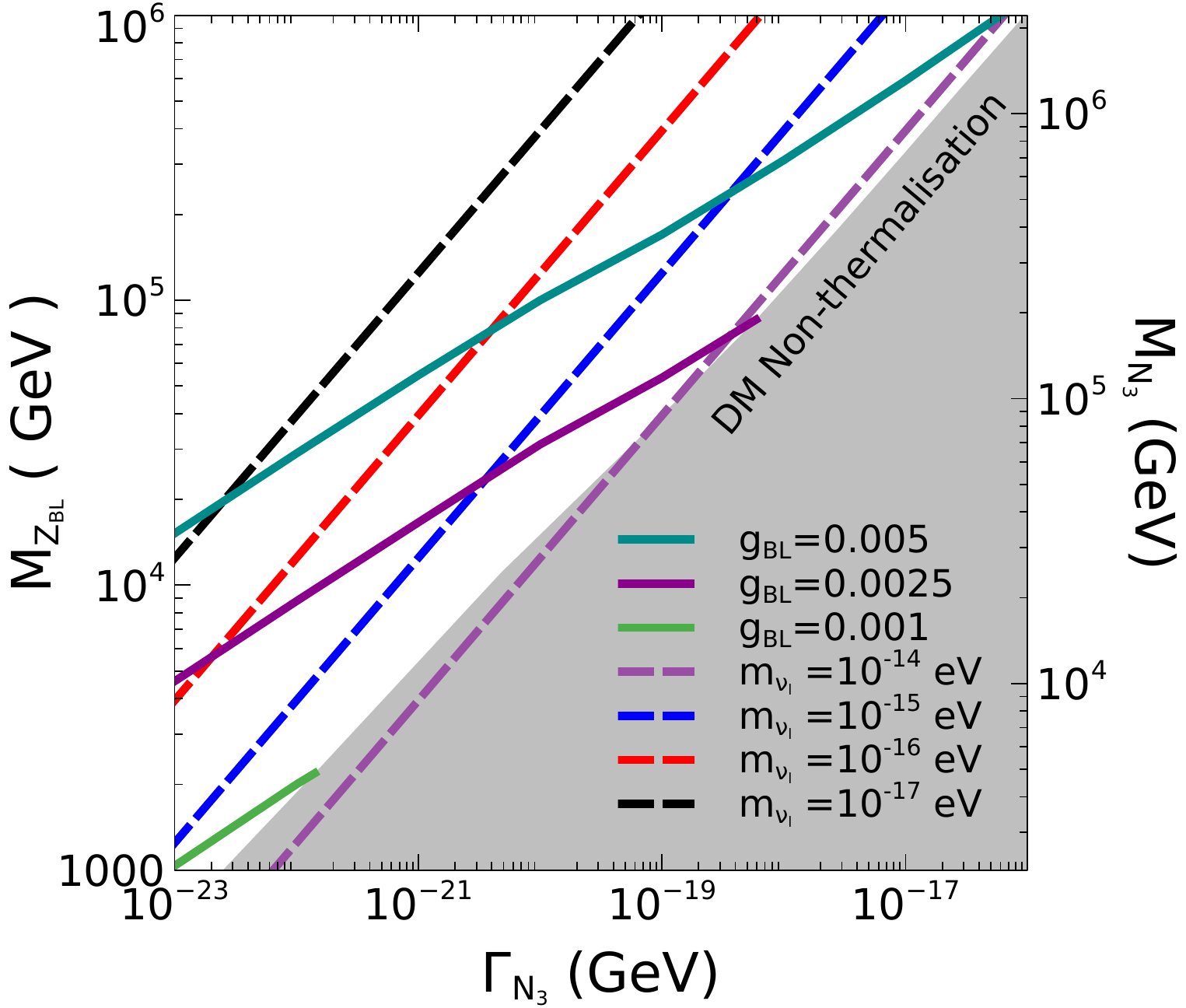}
\caption{Contours of lightest neutrino mass in $\Gamma_{N_3}-M_{Z_{BL}}(M_{N_2})$ plane. The DM relic satisfied contours are also shown for different $g_{BL}$ values considering $M_{\rm DM}=5$ keV.}
\label{fig:neumass}
\end{figure}

In Fig.\,\ref{fig:neumass}, we show different contours (dashed) for $m_{{\nu}_{ l}}=\{10^{-14},10^{-15},10^{-16},10^{-17}\}$ eV in the $\Gamma_{N_3}-M_{Z_{BL}}(M_{N_3})$ plane. We also show the relic density satisfied lines (solid) in the same plane for three different values of $g_{BL}$ considering $M_{\rm DM}=5$ keV.
The requirement of DM thermalisation in the early universe which rules out grey shaded region in Fig.\,\ref{fig:neumass}, restricts the lightest neutrino mass to be $m_{\nu_l}\lesssim \mathcal{O}(10^{-14})$ eV. 
We infer from this figure that the $B-L$ gauge coupling is correlated with the lightest active neutrino mass for a fixed set of $\Gamma_{N_3}$ and $M_{\rm DM}$. Any intersecting point between constant $g_{BL}$ and constant $m_{\nu_l}$ contours in Fig.\,\ref{fig:neumass} provides us suitable choices of the model parameters that simultaneously can yield correct DM relic with required entropy dilution and satisfy the neutrino oscillation data with definite prediction of lightest active neutrino mass. As an example, in table\,\ref{tab:tab2} we tabulate two  benchmark points that illustrate such interesting correlations.

\begin{table*}
\begin{center}
  \begin{tabular}{| c | c | c | c | c | c | c | }
    \hline
   &  $m_{\rm DM}$ & $g_{BL}$ & $M_{Z_{BL}}$\,(GeV) & $M_{N_3}$\,(GeV) & $\Gamma_{N_3}$\,(GeV) & $m_{\nu_{l}}$\,(eV) \\ \hline
    BP\,I & \multirow{2}{*}{5 keV} & 0.0025 & $7.7\times 10^{4}$ & $1.7\times 10^{5}$ & $4\times 10^{-19}$ & $\mathcal{O}(10^{-14})$ \\ \cline{1-1} \cline{3-7}
    BP\,II & & 0.005 & $8.2\times 10^{4}$ & $1.8\times 10^5$ & $4.7\times 10^{-21}$ & $\mathcal{O}(10^{-16})$\\ \hline
  \end{tabular}
\end{center}
\caption{Two representative benchmark points of our proposed set up that simultaneously provide correct DM relic and also uniquely predict the order of lightest active neutrino mass.}
\label{tab:tab2}
\end{table*}
Next, we check whether the proposed framework can accommodate successful baryogenesis via leptogenesis \cite{Fukugita:1986hr} in the presence of same amount of late time entropy production that dilutes the thermally overproduced DM relic. For this purpose we utilise the first benchmark point, tabulated in table\,\ref{tab:tab2}. The free parameters that remain unconstrained from the dynamics of inflation, phenomenology of dark matter and prediction of lightest neutrino mass are the mass scales $M_{N_1}, M_{N_2}$ and the rotation angle $\gamma$. The dependence of neutrino parameters data on $M_{N_1}, M_{N_2}$ can be absorbed in the respective Dirac Yukawa couplings. Note that, $N_3$ being feebly coupled to SM leptons, contributes negligibly to the production of lepton asymmetry, as compared to $N_1$ and $N_2$. The Boltzmann equations that govern the dynamics for $N_1$, $N_2$ decay and the yield of $B-L$ asymmetry are given by \cite{Iso:2010mv},
\begin{widetext}
\begin{align}
&aH\frac{d  n_1}{da}+3H n_{1}=-\gamma_{D_1}\left(\frac{n_1}{n_1^{\rm eq}}-1\right)-\left\{\left(\frac{n_1}{n_1^{\rm eq}}\right)^2-1\right\}\gamma_{Z_{BL}^1},\label{eq:lep1}\\
&aH\frac{d  n_2}{da}+3H n_{2}=-\gamma_{D_2}\left(\frac{n_2}{n_2^{\rm eq}}-1\right)-\left\{\left(\frac{n_2}{n_2^{\rm eq}}\right)^2-1\right\}\gamma_{Z_{BL}^2},\label{eq:lep2}\\
&aH\frac{d  n_{B-L}}{da}+3H n_{B-L}=-\sum_{j=1}^2\Bigg\{\frac{1}{2}\frac{n_{B-L}}{n_{L}^{\rm eq}}+\varepsilon_j\Bigg(\frac{n_{N_j}}{n_{N_j}^{\rm eq}}-1\Bigg)\Bigg\}\gamma_{D_j},
\label{eq:lep3}
\end{align}
\end{widetext}
where $n_i$($n_{B-L}$) are the number densities for $N_i$ ($B-L$ asymmetry) and $\varepsilon_i$ represents the lepton asymmetry parameter.  The standard convention to express the comoving abundance of $B-L$ asymmetry is given by $\frac{n_{B-L}}{s}$, with the entropy density defined as $s=\frac{2\pi^2}{45}g_{*s}T(a)^3$. The $\gamma_{D_i{}}$ includes the effects of decays and inverse decays while $\gamma_{Z_{BL}^i}$ takes into account the contribution from the $Z_{BL}$ mediated scatterings $ff\leftrightarrow N_iN_i$ (see \cite{Iso:2010mv} for the analytical expressions).  We do not include the other $2\rightarrow 2$ processes since their effects turn out to be sub-dominant in the present analysis \cite{Iso:2010mv,Okada:2012fs}. The temperature and the entropy density of the universe as function of scale factor are evaluated using Eq.(\ref{eq:Boltz}). Recall that earlier we have considered $Y_{N_{1,2}}\ll Y_{N_3}$. The value of $M_{N_3}$ is $\mathcal{O}(10^5)$ GeV in the first benchmark of table\,\ref{tab:tab2}.  For illustrative purpose, we consider a low scale leptogenesis and fix $M_{N_1}$ at 2 TeV and write $M_{N_2}=M_{N_1}+\Delta$. In that case the amount of lepton asymmetry essentially depends on $\Delta$ and $\gamma$ only, once the other parameters are fixed according to BP I in table\,\ref{tab:tab2} with $M_{N_1}=2$ TeV. Production of sufficient lepton asymmetry from such TeV scale RH neutrinos (violating the Davidson-Ibarra bound \cite{Davidson:2002qv}) is possible only in the resonant scenario \cite{Pilaftsis:2003gt}. In resonant case, {the CP asymmetry parameter $\varepsilon$ is maximised at $\Delta\sim \frac{\Gamma_i}{2}$. Although, this condition is not very strict and one can adjust $\Delta$ accordingly to obtain the appropriate value of CP asymmetry parameter $\varepsilon_i$ in order to yield observed baryon asymmetry depending on the strength of washout.}

\begin{figure*}[htb!]
\includegraphics[height=6.4cm,width=8.3cm]{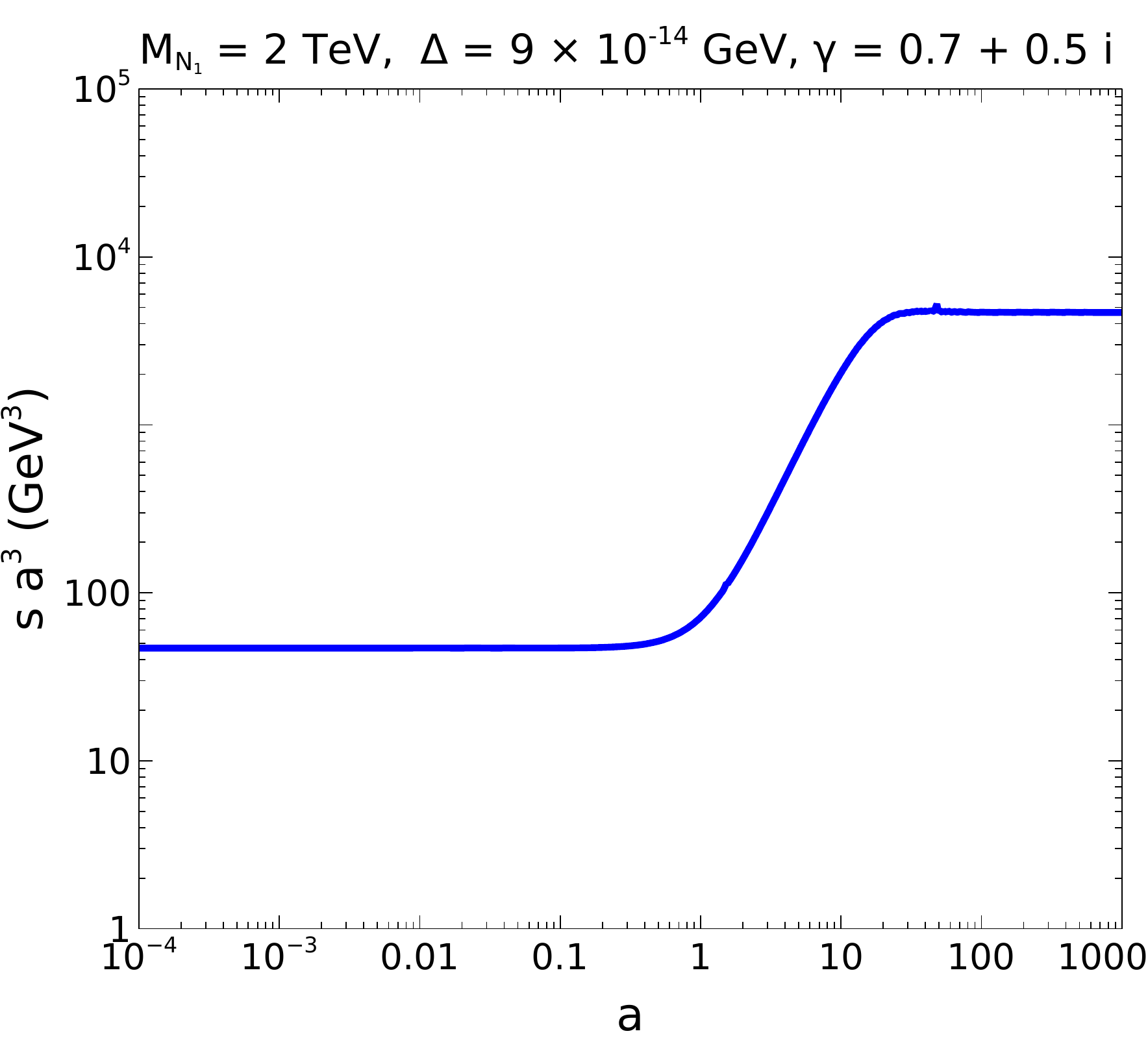}
\caption{Evolution of comoving entropy density as a function of scale factor for the BP I as tabulated in table\,\ref{tab:tab2}.}
\label{fig:entropyS}
\end{figure*}

Next, we look for a particular set of $\{\Delta,\gamma\}$ such that we obtain required order of baryon asymmetry (consistent with the observation), surviving the combined effect of washouts and late time entropy dilution. We have made the choice $\Delta=9\times 10^{-14}$ GeV and $\gamma=0.7+0.5i$ for the BP I in table\,\ref{tab:tab2}.
Below we note down the numerical estimate the neutrino Yuwkawa matrix as obtained by implementing the Casas-Ibarra parametrisation following Eq.\,(\ref{eq:CI}). The Yukawa structure for $N_1$ and $N_2$ turns out to be (considering the mass of the lightest active neutrino vanishing),
\begin{widetext}
\begin{align} 
Y_D^{3\times 2}=\left(
\begin{array}{ccc}
 1.67 \times 10^{-7}\, +5.52 \times 10^{-7}\,i& 9 \times 10^{-7}\, +1.48 \times 10^{-9}\,i \\
 -3.42 \times 10^{-7}\, - 4.32 \times 10^{-7}\,i & -1.13 \times 10^{-6}\,  + 2.12 \times 10^{-7}\,i\\
 1.42 \times 10^{-6}\, + 3.78 \times 10^{-7}\,i & 8.19 \times 10^{-7}\, - 6.54 \times 10^{-7}\,i
\end{array}
\right)\,.
\end{align}
\end{widetext}
This particular Yukawa structure yields following outputs
\begin{align}
&\Gamma_{N_1}=  2.2\times10^{-10},~~~~\Gamma_{N_2}=2.6\times10^{-10} \\
&\varepsilon_1=0.00048, ~~~~\varepsilon_2=0.00056.
\end{align}
 \noindent We also find $\Delta=0.0004\times \Gamma_{N_1}=0.00034\times\Gamma_{N_2}$. It is pertinent to affirm that the neutrino oscillation data are automatically satisfied since we have used the best-fit values for $\Delta m_{21}^2$, $\Delta m_{31}^2$ and neutrino mixing angles in the Casas-Ibarra parametrisation as earlier mentioned considering vanishing lightest active neutrino mass. We have solved Eqs.(\ref{eq:lep1})-(\ref{eq:lep3}) assuming $N_1$ and $N_2$ to follow the equilibrium distribution initially with the temperature same as the one of SM thermal bath. In Fig.\,\ref{fig:entropyS}, we plot the variation of comoving entropy density with scale factor for the BP I as tabulated in table\,\ref{tab:tab2} clearly showing a late time enhancement of the entropy due to the decay of long-lived $N_3$ into thermal plasma.

\begin{figure*}[htb!]
\includegraphics[height=6.4cm,width=8.3cm]{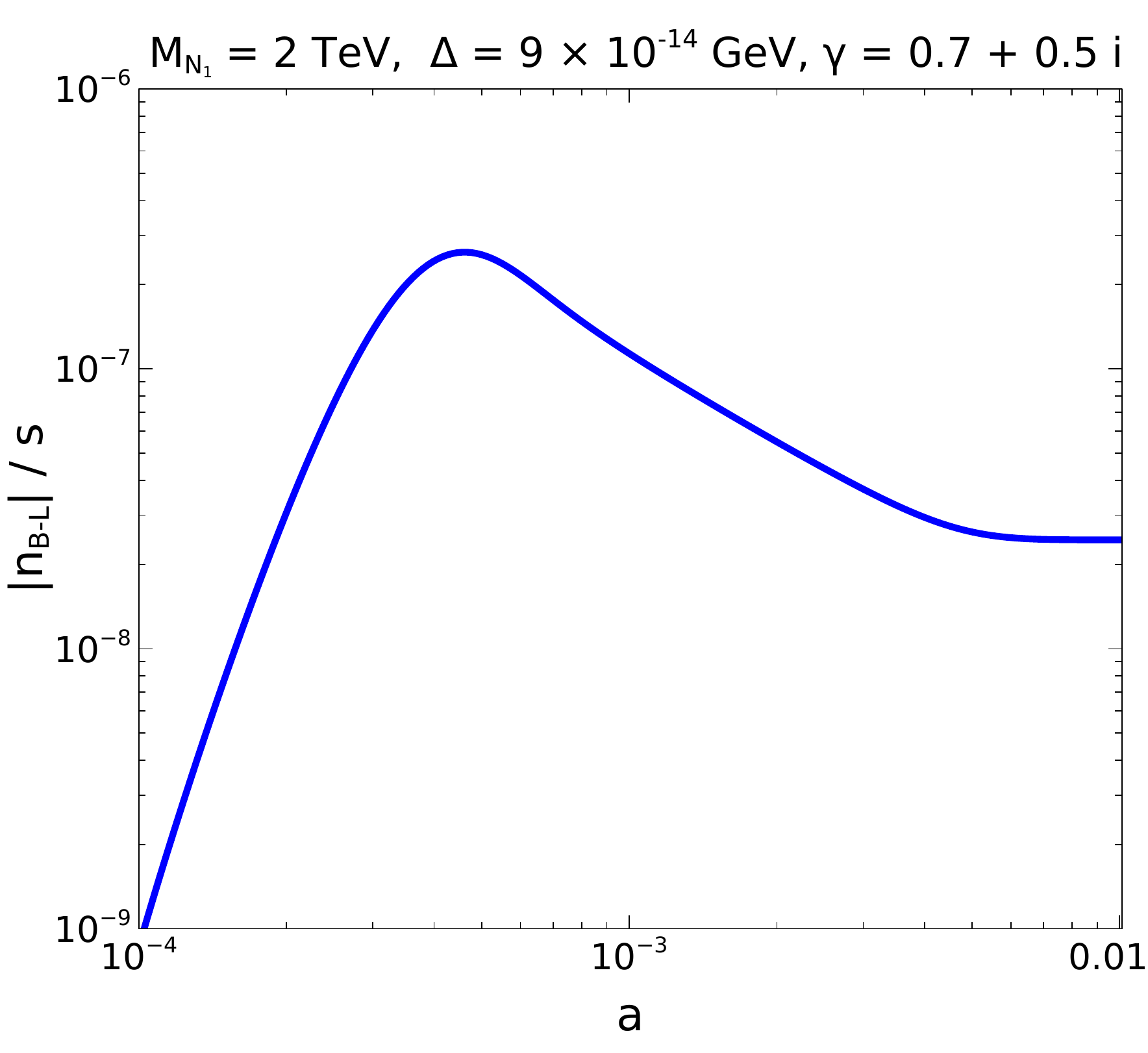}
\includegraphics[height=6.4cm,width=8.3cm]{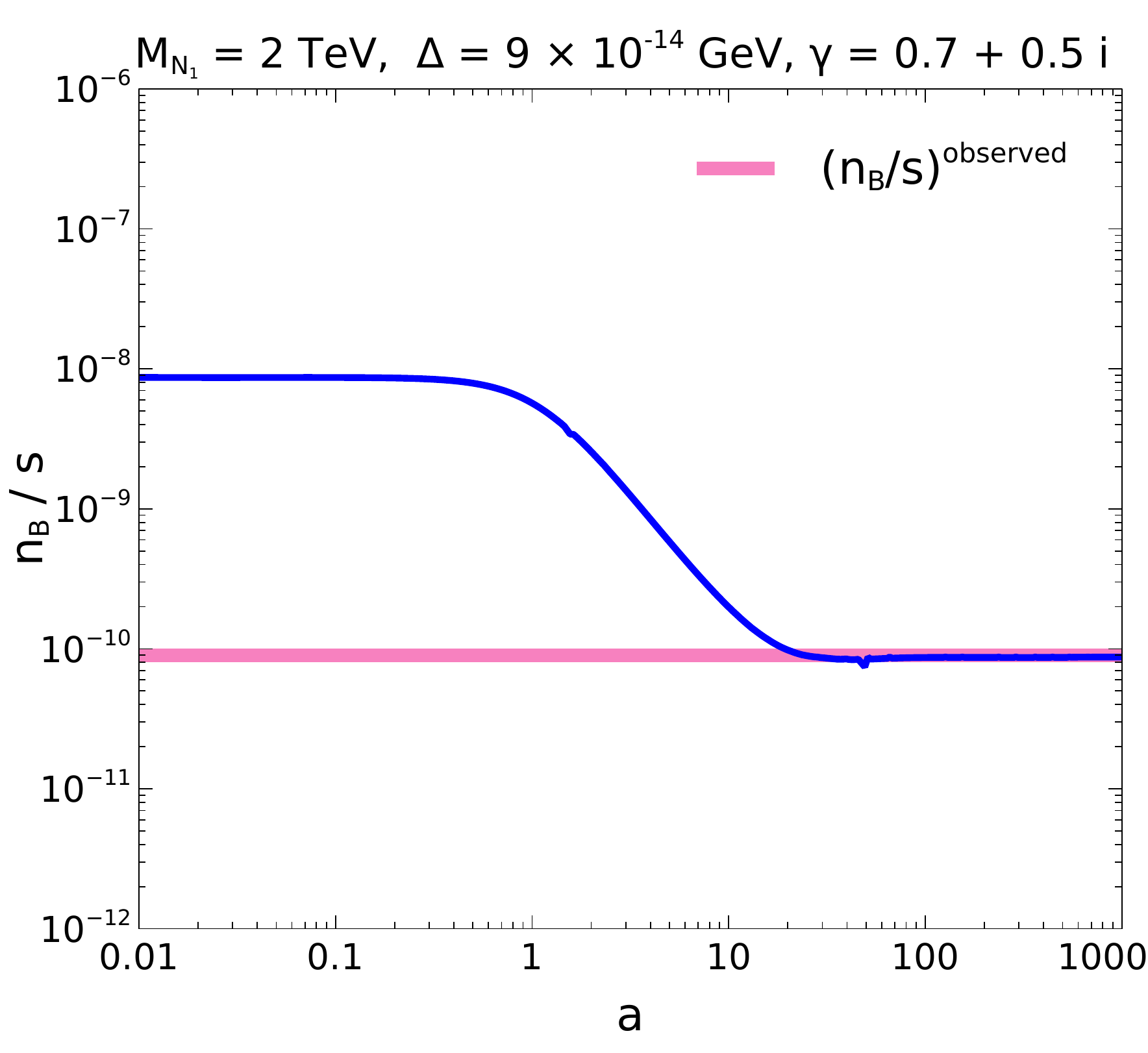}
\caption{Left panel: Evolution of comoving $B-L$ asymmetry as function of scale factor for the benchmark point I as tabulated in table\,\ref{tab:tab2} till $T\sim 100$ GeV. Right panel: The evolution of comoving baryon asymmetry from $T\sim 100$ GeV till the present epoch. The choice of $\Delta$ and $\gamma$ have been made such that we attain observed value of baryon asymmetry $\sim  9\times10^{-11}$\cite{Planck:2018vyg} (as indicated by the horizontal red line) of the universe considering a TeV scale $N_1$.}
\label{fig:lepS}
\end{figure*}

 The generated lepton asymmetry gets converted into baryon asymmetry of the universe via standard sphaleron process prior to the sphaleron decoupling at $T\sim 100$ GeV. This conversion can be quantified as,
\begin{align}
\frac{n_B}{s}=-\frac{28}{79}\frac{n_{B-L}}{s}\label{eq:conBL}
\end{align}
In left panel of Fig.\,\ref{fig:lepS}, we show the evolution of  $\frac{|n_{B-L}|}{s}$ as function of scale factor considering the set  $\{\Delta=9\times 10^{-14}\,{\rm GeV},\gamma=0.7+0.5 i\}$ till $a\sim 0.01$ which approximately implies $T\sim 100$ GeV. At this temperature, the conversion of lepton asymmetry to baryon asymmetry ends since the $SU(2)_L$ sphaleron processes get switched off. In the right panel of Fig.\,\ref{fig:lepS}, we have shown the evolution pattern for baryon asymmetry from $T\sim 100$ GeV till the present epoch with the final value matching with the observed baryon asymmetry.

 We see that initially $\frac{|n_{B-L}|}{s}$ gets produced from the out of equilibrium decay of $N_1$ and $N_2$. Once the production ends, $\frac{|n_{B-L}|}{s}$ decreases to some extent due to the wash-out effects like inverse decay and then freezes to a particular amount, resulting in the plateau region till $a\sim 0.01$. At this point with $T\sim 100$ GeV, we use Eq.(\ref{eq:conBL}) to obtain the $\frac{n_B}{s}$ and then find out its evolution as function of scale factor till present epoch. From Fig.\,\ref{fig:entropyS}, we see a late time enhancement of the entropy due to long-lived $N_3$ decay into radiation. This results in dilution of the $\frac{n_B}{s}$ from its value at $a\sim 0.01$ and we finally obtain the remnant amount of $\frac{n_B}{s}\sim$ ($9\times 10^{-11}$) which overlaps with the observed baryon asymmetry of the universe \cite{Planck:2018vyg}. Thus the present framework indeed offers the possibility to realise successful baryogenesis via leptogenesis satisfying inflationary bounds on model parameters and WDM relic abundance with significant late time entropy dilution.

As mentioned earlier, keV scale warm dark matter can alleviate some of the small scale structure issues of cold dark matter paradigm. Due to larger free-streaming length, such WDM scenarios can give rise to different structure formation rates which can be probed by several galaxy survey experiments. For example, the authors of \cite{Newton:2020cog} constrained thermal WDM mass around keV scale by comparing the predictions of DM sub-structures in Milky Way with the estimates of total satellite galaxy population. Similar bounds on thermal WDM mass were also derived in \cite{Banik:2019smi} by using constraints from stellar streams. It should be noted that the free streaming length of WDM can get slightly modified due to the late time entropy dilution in our setup but expected to keep the lower bound on WDM mass in the order of $\mathcal{O}(1)$\,keV.


 \section{Conclusion}
 \label{sec:conclude}
 We have proposed a simple extension of the minimal gauged $B-L$ model in order to realise the possibility of realising cosmic inflation, keV scale warm dark matter and light neutrino mass and mixing simultaneously. A compatible picture of cosmic inflation poses strong constraints on the minimal $B-L$ model parameters. After performing a renormalisation group improved analysis of inflationary dynamics leading to agreement of predicted inflationary parameters with the latest Planck+BICEP/Keck results of 2021, we obtain the parameter space of the proposed model by incorporating all other relevant constraints including the collider, BBN bounds. On the other hand, a thermal DM candidate of keV scale (a gauge singlet vector like fermion) leads to overproduction as it freezes out while being relativistic like active neutrinos. We invoke the late entropy dilution mechanism from a long-lived heavier RHN ($N_3$, to be specific) in order to bring the overproduced keV scale DM abundance within Planck 2018 limits. The requirement of sufficient entropy dilution demands $N_3$ to freeze-out in the early universe leading to an intermediate matter dominated epoch. The thermalisation temperature of DM and the decaying particle in the early universe as well as the efficiency of the entropy dilution purely depend on their interactions regulated by the $B-L$ model parameters. Due to the minimal nature of the model and the involvement of cosmic inflation, we indeed find a small parameter parameter space in agreement with all these requirements. In particular, we have observed that to achieve correct relic abundance of dark matter, it is preferable to have heavier $Z_{BL}$ ($\gtrsim \mathcal{O}$(1) TeV ) with larger lifetime of $N_3$. Additionally, we also predict the lightest active neutrino mass to be $m_{\nu_l}\lesssim 10^{-14}$ eV considering a keV order WDM mass and satisfying the neutrino oscillation data. This prediction is also dependent on the $B-L$ model parameters with higher $g_{BL}$ leading to lowering of $m_{\nu_l}$ further for a constant lifetime of $N_3$. Such tiny values of lightest active neutrino mass, as predicted by the model, will keep the effective neutrino mass much out of reach from ongoing tritium beta decay experiments like KATRIN \cite{KATRIN:2019yun}. {In addition, near
future observation of neutrinoless double beta decay \cite{Dolinski:2019nrj} can also falsify our scenario, particularly for normal ordering of light neutrinos.}

 Finally, we check the possibility of thermal leptogenesis from the out-of-equilibrium decay of other two RHNs namely, $N_1$ and $N_2$ in the presence of long lived $N_3$. We have found that sufficient amount of lepton asymmetry can be obtained in the resonant regime which can survive the late time injection of large entropy, and can get converted into the observed baryon asymmetry via electroweak sphalerons. In addition to the predictive nature as far as model parameters are concerned, warm dark matter of keV scale also has interesting astrophysical implications in terms of small scale structure formation. By allowing a tiny mixing of such WDM with active neutrinos will also give rise to the the possibility of monochromatic photon from its radiative decay leading to tantalizing indirect detection prospects.

\section*{Acknowledgement}
 AKS is supported by NPDF grant PDF/2020/000797 from Science and Engineering Research Board, Government of India.
 

\begin{thebibliography}{10}

\bibitem{Jungman:1995df}
G.~Jungman, M.~Kamionkowski, and K.~Griest, {\it {Supersymmetric dark matter}},
   {\em Phys. Rept.} {\bf 267} (1996) 195--373,
  [\href{http://arxiv.org/abs/hep-ph/9506380}{{\tt hep-ph/9506380}}].

\bibitem{Feng:2010gw}
J.~L. Feng, {\it {Dark Matter Candidates from Particle Physics and Methods of
  Detection}},  {\em Ann. Rev. Astron. Astrophys.} {\bf 48} (2010) 495--545,
  [\href{http://arxiv.org/abs/1003.0904}{{\tt arXiv:1003.0904}}].

\bibitem{Bertone:2004pz}
G.~Bertone, D.~Hooper, and J.~Silk, {\it {Particle dark matter: Evidence,
  candidates and constraints}},  {\em Phys. Rept.} {\bf 405} (2005) 279--390,
  [\href{http://arxiv.org/abs/hep-ph/0404175}{{\tt hep-ph/0404175}}].

\bibitem{ParticleDataGroup:2020ssz}
{\bf Particle Data Group} Collaboration, P.~A. Zyla et~al., {\it {Review of
  Particle Physics}},  {\em PTEP} {\bf 2020} (2020), no.~8 083C01.

\bibitem{Planck:2018vyg}
{\bf Planck} Collaboration, N.~Aghanim et~al., {\it {Planck 2018 results. VI.
  Cosmological parameters}},  {\em Astron. Astrophys.} {\bf 641} (2020) A6,
  [\href{http://arxiv.org/abs/1807.06209}{{\tt arXiv:1807.06209}}]. [Erratum:
  Astron.Astrophys. 652, C4 (2021)].

\bibitem{Kolb:1990vq}
E.~W. Kolb and M.~S. Turner, {\em {The Early Universe}}, vol.~69.
\newblock 1990.

\bibitem{LUX:2016ggv}
{\bf LUX} Collaboration, D.~S. Akerib et~al., {\it {Results from a search for
  dark matter in the complete LUX exposure}},  {\em Phys. Rev. Lett.} {\bf 118}
  (2017), no.~2 021303, [\href{http://arxiv.org/abs/1608.07648}{{\tt
  arXiv:1608.07648}}].

\bibitem{PandaX-4T:2021bab}
{\bf PandaX-4T} Collaboration, Y.~Meng et~al., {\it {Dark Matter Search Results
  from the PandaX-4T Commissioning Run}},  {\em Phys. Rev. Lett.} {\bf 127}
  (2021), no.~26 261802, [\href{http://arxiv.org/abs/2107.13438}{{\tt
  arXiv:2107.13438}}].

\bibitem{XENON:2018voc}
{\bf XENON} Collaboration, E.~Aprile et~al., {\it {Dark Matter Search Results
  from a One Ton-Year Exposure of XENON1T}},  {\em Phys. Rev. Lett.} {\bf 121}
  (2018), no.~11 111302, [\href{http://arxiv.org/abs/1805.12562}{{\tt
  arXiv:1805.12562}}].

\bibitem{Kahlhoefer:2017dnp}
F.~Kahlhoefer, {\it {Review of LHC Dark Matter Searches}},  {\em Int. J. Mod.
  Phys. A} {\bf 32} (2017), no.~13 1730006,
  [\href{http://arxiv.org/abs/1702.02430}{{\tt arXiv:1702.02430}}].

\bibitem{Penning:2017tmb}
B.~Penning, {\it {The pursuit of dark matter at colliders\textemdash{}an
  overview}},  {\em J. Phys. G} {\bf 45} (2018), no.~6 063001,
  [\href{http://arxiv.org/abs/1712.01391}{{\tt arXiv:1712.01391}}].

\bibitem{Boyarsky:2008xj}
A.~Boyarsky, J.~Lesgourgues, O.~Ruchayskiy, and M.~Viel, {\it {Lyman-alpha
  constraints on warm and on warm-plus-cold dark matter models}},  {\em JCAP}
  {\bf 05} (2009) 012, [\href{http://arxiv.org/abs/0812.0010}{{\tt
  arXiv:0812.0010}}].

\bibitem{Merle:2013wta}
A.~Merle, V.~Niro, and D.~Schmidt, {\it {New Production Mechanism for keV
  Sterile Neutrino Dark Matter by Decays of Frozen-In Scalars}},  {\em JCAP}
  {\bf 03} (2014) 028, [\href{http://arxiv.org/abs/1306.3996}{{\tt
  arXiv:1306.3996}}].

\bibitem{Drewes:2016upu}
M.~Drewes et~al., {\it {A White Paper on keV Sterile Neutrino Dark Matter}},
  {\em JCAP} {\bf 01} (2017) 025, [\href{http://arxiv.org/abs/1602.04816}{{\tt
  arXiv:1602.04816}}].

\bibitem{Gorbunov:2008ka}
D.~Gorbunov, A.~Khmelnitsky, and V.~Rubakov, {\it {Constraining sterile
  neutrino dark matter by phase-space density observations}},  {\em JCAP} {\bf
  10} (2008) 041, [\href{http://arxiv.org/abs/0808.3910}{{\tt
  arXiv:0808.3910}}].

\bibitem{Boyarsky:2008ju}
A.~Boyarsky, O.~Ruchayskiy, and D.~Iakubovskyi, {\it {A Lower bound on the mass
  of Dark Matter particles}},  {\em JCAP} {\bf 03} (2009) 005,
  [\href{http://arxiv.org/abs/0808.3902}{{\tt arXiv:0808.3902}}].

\bibitem{Seljak:2006qw}
U.~Seljak, A.~Makarov, P.~McDonald, and H.~Trac, {\it {Can sterile neutrinos be
  the dark matter?}},  {\em Phys. Rev. Lett.} {\bf 97} (2006) 191303,
  [\href{http://arxiv.org/abs/astro-ph/0602430}{{\tt astro-ph/0602430}}].

\bibitem{Bulbul:2014sua}
E.~Bulbul, M.~Markevitch, A.~Foster, R.~K. Smith, M.~Loewenstein, and S.~W.
  Randall, {\it {Detection of An Unidentified Emission Line in the Stacked
  X-ray spectrum of Galaxy Clusters}},  {\em Astrophys. J.} {\bf 789} (2014)
  13, [\href{http://arxiv.org/abs/1402.2301}{{\tt arXiv:1402.2301}}].

\bibitem{Boyarsky:2014jta}
A.~Boyarsky, O.~Ruchayskiy, D.~Iakubovskyi, and J.~Franse, {\it {Unidentified
  Line in X-Ray Spectra of the Andromeda Galaxy and Perseus Galaxy Cluster}},
  {\em Phys. Rev. Lett.} {\bf 113} (2014) 251301,
  [\href{http://arxiv.org/abs/1402.4119}{{\tt arXiv:1402.4119}}].

\bibitem{Bullock:2017xww}
J.~S. Bullock and M.~Boylan-Kolchin, {\it {Small-Scale Challenges to the
  $\Lambda$CDM Paradigm}},  {\em Ann. Rev. Astron. Astrophys.} {\bf 55} (2017)
  343--387, [\href{http://arxiv.org/abs/1707.04256}{{\tt arXiv:1707.04256}}].

\bibitem{Guth:1980zm}
A.~H. Guth, {\it {The Inflationary Universe: A Possible Solution to the Horizon
  and Flatness Problems}},  {\em Phys. Rev. D} {\bf 23} (1981) 347--356.

\bibitem{Starobinsky:1980te}
A.~A. Starobinsky, {\it {A New Type of Isotropic Cosmological Models Without
  Singularity}},  {\em Phys. Lett. B} {\bf 91} (1980) 99--102.

\bibitem{Linde:1981mu}
A.~D. Linde, {\it {A New Inflationary Universe Scenario: A Possible Solution of
  the Horizon, Flatness, Homogeneity, Isotropy and Primordial Monopole
  Problems}},  {\em Phys. Lett. B} {\bf 108} (1982) 389--393.

\bibitem{Bezrukov:2007ep}
F.~L. Bezrukov and M.~Shaposhnikov, {\it {The Standard Model Higgs boson as the
  inflaton}},  {\em Phys. Lett. B} {\bf 659} (2008) 703--706,
  [\href{http://arxiv.org/abs/0710.3755}{{\tt arXiv:0710.3755}}].

\bibitem{Lerner:2009na}
R.~N. Lerner and J.~McDonald, {\it {Higgs Inflation and Naturalness}},  {\em
  JCAP} {\bf 04} (2010) 015, [\href{http://arxiv.org/abs/0912.5463}{{\tt
  arXiv:0912.5463}}].

\bibitem{Okada:2011en}
N.~Okada, M.~U. Rehman, and Q.~Shafi, {\it {Non-Minimal B-L Inflation with
  Observable Gravity Waves}},  {\em Phys. Lett. B} {\bf 701} (2011) 520--525,
  [\href{http://arxiv.org/abs/1102.4747}{{\tt arXiv:1102.4747}}].

\bibitem{Okada:2015lia}
N.~Okada and D.~Raut, {\it {Running non-minimal inflation with stabilized
  inflaton potential}},  {\em Eur. Phys. J. C} {\bf 77} (2017), no.~4 247,
  [\href{http://arxiv.org/abs/1509.04439}{{\tt arXiv:1509.04439}}].

\bibitem{Borah:2020wyc}
D.~Borah, S.~Jyoti~Das, and A.~K. Saha, {\it {Cosmic inflation in minimal
  $U(1)_{B-L}$ model: implications for (non) thermal dark matter and
  leptogenesis}},  {\em Eur. Phys. J. C} {\bf 81} (2021), no.~2 169,
  [\href{http://arxiv.org/abs/2005.11328}{{\tt arXiv:2005.11328}}].

\bibitem{BICEPKeck:2021gln}
{\bf BICEP, Keck} Collaboration, P.~A.~R. Ade et~al., {\it {Improved
  Constraints on Primordial Gravitational Waves using Planck, WMAP, and
  BICEP/Keck Observations through the 2018 Observing Season}},  {\em Phys. Rev.
  Lett.} {\bf 127} (2021), no.~15 151301,
  [\href{http://arxiv.org/abs/2110.00483}{{\tt arXiv:2110.00483}}].

\bibitem{Nemevsek:2012cd}
M.~Nemevsek, G.~Senjanovic, and Y.~Zhang, {\it {Warm Dark Matter in Low Scale
  Left-Right Theory}},  {\em JCAP} {\bf 07} (2012) 006,
  [\href{http://arxiv.org/abs/1205.0844}{{\tt arXiv:1205.0844}}].

\bibitem{Bezrukov:2009th}
F.~Bezrukov, H.~Hettmansperger, and M.~Lindner, {\it {keV sterile neutrino Dark
  Matter in gauge extensions of the Standard Model}},  {\em Phys. Rev. D} {\bf
  81} (2010) 085032, [\href{http://arxiv.org/abs/0912.4415}{{\tt
  arXiv:0912.4415}}].

\bibitem{Borah:2017hgt}
D.~Borah and A.~Dasgupta, {\it {Left\textendash{}right symmetric models with a
  mixture of keV\textendash{}TeV dark matter}},  {\em J. Phys. G} {\bf 46}
  (2019), no.~10 105004, [\href{http://arxiv.org/abs/1710.06170}{{\tt
  arXiv:1710.06170}}].

\bibitem{Dror:2020jzy}
J.~A. Dror, D.~Dunsky, L.~J. Hall, and K.~Harigaya, {\it {Sterile Neutrino Dark
  Matter in Left-Right Theories}},  {\em JHEP} {\bf 07} (2020) 168,
  [\href{http://arxiv.org/abs/2004.09511}{{\tt arXiv:2004.09511}}].

\bibitem{Dutra:2021lto}
M.~Dutra, V.~Oliveira, C.~A. de~S.~Pires, and F.~S. Queiroz, {\it {A model for
  mixed warm and hot right-handed neutrino dark matter}},  {\em JHEP} {\bf 10}
  (2021) 005, [\href{http://arxiv.org/abs/2104.14542}{{\tt arXiv:2104.14542}}].

\bibitem{Arcadi:2021doo}
G.~Arcadi, J.~P. Neto, F.~S. Queiroz, and C.~Siqueira, {\it {Roads for
  right-handed neutrino dark matter: Fast expansion, standard freeze-out, and
  early matter domination}},  {\em Phys. Rev. D} {\bf 105} (2022), no.~3
  035016, [\href{http://arxiv.org/abs/2108.11398}{{\tt arXiv:2108.11398}}].

\bibitem{Davidson:1978pm}
A.~Davidson, {\it {$B-L$ as the fourth color within an $ SU(2)_L \times U(1)_R \times U(1)$  model}},  {\em Phys. Rev. D}
  {\bf 20} (1979) 776.

\bibitem{Mohapatra:1980qe}
R.~N. Mohapatra and R.~E. Marshak, {\it {Local B-L Symmetry of Electroweak
  Interactions, Majorana Neutrinos and Neutron Oscillations}},  {\em Phys. Rev.
  Lett.} {\bf 44} (1980) 1316--1319. [Erratum: Phys.Rev.Lett. 44, 1643 (1980)].

\bibitem{Marshak:1979fm}
R.~E. Marshak and R.~N. Mohapatra, {\it {Quark - Lepton Symmetry and B-L as the
  U(1) Generator of the Electroweak Symmetry Group}},  {\em Phys. Lett. B} {\bf
  91} (1980) 222--224.

\bibitem{Masiero:1982fi}
A.~Masiero, J.~F. Nieves, and T.~Yanagida, {\it {$B^-$l Violating Proton Decay
  and Late Cosmological Baryon Production}},  {\em Phys. Lett. B} {\bf 116}
  (1982) 11--15.

\bibitem{Mohapatra:1982xz}
R.~N. Mohapatra and G.~Senjanovic, {\it {Spontaneous Breaking of Global $B^-$l
  Symmetry and Matter - Antimatter Oscillations in Grand Unified Theories}},
  {\em Phys. Rev. D} {\bf 27} (1983) 254.

\bibitem{Buchmuller:1991ce}
W.~Buchmuller, C.~Greub, and P.~Minkowski, {\it {Neutrino masses, neutral
  vector bosons and the scale of B-L breaking}},  {\em Phys. Lett. B} {\bf 267}
  (1991) 395--399.

\bibitem{Mambrini:2011dw}
Y.~Mambrini, {\it {The ZZ' kinetic mixing in the light of the recent direct and
  indirect dark matter searches}},  {\em JCAP} {\bf 07} (2011) 009,
  [\href{http://arxiv.org/abs/1104.4799}{{\tt arXiv:1104.4799}}].

\bibitem{Carena:2004xs}
M.~Carena, A.~Daleo, B.~A. Dobrescu, and T.~M.~P. Tait, {\it {$Z^\prime$ gauge
  bosons at the Tevatron}},  {\em Phys. Rev. D} {\bf 70} (2004) 093009,
  [\href{http://arxiv.org/abs/hep-ph/0408098}{{\tt hep-ph/0408098}}].

\bibitem{Cacciapaglia:2006pk}
G.~Cacciapaglia, C.~Csaki, G.~Marandella, and A.~Strumia, {\it {The Minimal Set
  of Electroweak Precision Parameters}},  {\em Phys. Rev. D} {\bf 74} (2006)
  033011, [\href{http://arxiv.org/abs/hep-ph/0604111}{{\tt hep-ph/0604111}}].

\bibitem{Aad:2019fac}
{\bf ATLAS} Collaboration, G.~Aad et~al., {\it {Search for high-mass dilepton
  resonances using 139 fb$^{-1}$ of $pp$ collision data collected at
  $\sqrt{s}=$13 TeV with the ATLAS detector}},  {\em Phys. Lett. B} {\bf 796}
  (2019) 68--87, [\href{http://arxiv.org/abs/1903.06248}{{\tt
  arXiv:1903.06248}}].

\bibitem{Sirunyan:2018exx}
{\bf CMS} Collaboration, A.~M. Sirunyan et~al., {\it {Search for high-mass
  resonances in dilepton final states in proton-proton collisions at
  $\sqrt{s}=$ 13 TeV}},  {\em JHEP} {\bf 06} (2018) 120,
  [\href{http://arxiv.org/abs/1803.06292}{{\tt arXiv:1803.06292}}].

\bibitem{Robens:2015gla}
T.~Robens and T.~Stefaniak, {\it {Status of the Higgs Singlet Extension of the
  Standard Model after LHC Run 1}},  {\em Eur. Phys. J. C} {\bf 75} (2015) 104,
  [\href{http://arxiv.org/abs/1501.02234}{{\tt arXiv:1501.02234}}].

\bibitem{Chalons:2016jeu}
G.~Chalons, D.~Lopez-Val, T.~Robens, and T.~Stefaniak, {\it {The Higgs singlet
  extension at LHC Run 2}},  {\em PoS} {\bf ICHEP2016} (2016) 1180,
  [\href{http://arxiv.org/abs/1611.03007}{{\tt arXiv:1611.03007}}].

\bibitem{Lopez-Val:2014jva}
D.~L\'opez-Val and T.~Robens, {\it {\ensuremath{\Delta}r and the W-boson mass
  in the singlet extension of the standard model}},  {\em Phys. Rev. D} {\bf
  90} (2014) 114018, [\href{http://arxiv.org/abs/1406.1043}{{\tt
  arXiv:1406.1043}}].

\bibitem{Khachatryan:2015cwa}
{\bf CMS} Collaboration, V.~Khachatryan et~al., {\it {Search for a Higgs boson
  in the mass range from 145 to 1000 GeV decaying to a pair of W or Z bosons}},
   {\em JHEP} {\bf 10} (2015) 144, [\href{http://arxiv.org/abs/1504.00936}{{\tt
  arXiv:1504.00936}}].

\bibitem{Strassler:2006ri}
M.~J. Strassler and K.~M. Zurek, {\it {Discovering the Higgs through
  highly-displaced vertices}},  {\em Phys. Lett. B} {\bf 661} (2008) 263--267,
  [\href{http://arxiv.org/abs/hep-ph/0605193}{{\tt hep-ph/0605193}}].

\bibitem{ATLAS:2020cjb}
{\bf ATLAS} Collaboration, {\it {Search for invisible Higgs boson decays with
  vector boson fusion signatures with the ATLAS detector using an integrated
  luminosity of 139 fb$^{-1}$}}, .

\bibitem{Capozziello:1996xg}
S.~Capozziello, R.~de~Ritis, and A.~A. Marino, {\it {Some aspects of the
  cosmological conformal equivalence between 'Jordan frame' and 'Einstein
  frame'}},  {\em Class. Quant. Grav.} {\bf 14} (1997) 3243--3258,
  [\href{http://arxiv.org/abs/gr-qc/9612053}{{\tt gr-qc/9612053}}].

\bibitem{Kaiser:2010ps}
D.~I. Kaiser, {\it {Conformal Transformations with Multiple Scalar Fields}},
  {\em Phys. Rev. D} {\bf 81} (2010) 084044,
  [\href{http://arxiv.org/abs/1003.1159}{{\tt arXiv:1003.1159}}].

\bibitem{Okada:2010jf}
N.~Okada, M.~U. Rehman, and Q.~Shafi, {\it {Tensor to Scalar Ratio in
  Non-Minimal $\phi^4$ Inflation}},  {\em Phys. Rev. D} {\bf 82} (2010) 043502,
  [\href{http://arxiv.org/abs/1005.5161}{{\tt arXiv:1005.5161}}].

\bibitem{Linde:2007fr}
A.~D. Linde, {\it {Inflationary Cosmology}},  {\em Lect. Notes Phys.} {\bf 738}
  (2008) 1--54, [\href{http://arxiv.org/abs/0705.0164}{{\tt arXiv:0705.0164}}].

\bibitem{Scherrer:1984fd}
R.~J. Scherrer and M.~S. Turner, {\it {Decaying Particles Do Not Heat Up the
  Universe}},  {\em Phys. Rev. D} {\bf 31} (1985) 681.

\bibitem{Arias:2019uol}
P.~Arias, N.~Bernal, A.~Herrera, and C.~Maldonado, {\it {Reconstructing
  Non-standard Cosmologies with Dark Matter}},  {\em JCAP} {\bf 10} (2019) 047,
  [\href{http://arxiv.org/abs/1906.04183}{{\tt arXiv:1906.04183}}].

\bibitem{Biswas:2018iny}
A.~Biswas, D.~Borah, and D.~Nanda, {\it {keV Neutrino Dark Matter in a Fast
  Expanding Universe}},  {\em Phys. Lett. B} {\bf 786} (2018) 364--372,
  [\href{http://arxiv.org/abs/1809.03519}{{\tt arXiv:1809.03519}}].

\bibitem{Allahverdi:2021grt}
R.~Allahverdi and J.~K. Osi\'nski, {\it {Early matter domination from
  long-lived particles in the visible sector}},  {\em Phys. Rev. D} {\bf 105}
  (2022), no.~2 023502, [\href{http://arxiv.org/abs/2108.13136}{{\tt
  arXiv:2108.13136}}].

\bibitem{Bhatia:2020itt}
D.~Bhatia and S.~Mukhopadhyay, {\it {Unitarity limits on thermal dark matter in
  (non-)standard cosmologies}},  {\em JHEP} {\bf 03} (2021) 133,
  [\href{http://arxiv.org/abs/2010.09762}{{\tt arXiv:2010.09762}}].

\bibitem{Escudero:2018mvt}
M.~Escudero, {\it {Neutrino decoupling beyond the Standard Model: CMB
  constraints on the Dark Matter mass with a fast and precise $N_{\rm eff}$
  evaluation}},  {\em JCAP} {\bf 02} (2019) 007,
  [\href{http://arxiv.org/abs/1812.05605}{{\tt arXiv:1812.05605}}].

\bibitem{Casas:2001sr}
J.~A. Casas and A.~Ibarra, {\it {Oscillating neutrinos and $\mu \to e,
  \gamma$}},  {\em Nucl. Phys. B} {\bf 618} (2001) 171--204,
  [\href{http://arxiv.org/abs/hep-ph/0103065}{{\tt hep-ph/0103065}}].

\bibitem{Ibarra:2003up}
A.~Ibarra and G.~G. Ross, {\it {Neutrino phenomenology: The Case of two
  right-handed neutrinos}},  {\em Phys. Lett. B} {\bf 591} (2004) 285--296,
  [\href{http://arxiv.org/abs/hep-ph/0312138}{{\tt hep-ph/0312138}}].

\bibitem{Fukugita:1986hr}
M.~Fukugita and T.~Yanagida, {\it {Baryogenesis Without Grand Unification}},
  {\em Phys. Lett. B} {\bf 174} (1986) 45--47.

\bibitem{Iso:2010mv}
S.~Iso, N.~Okada, and Y.~Orikasa, {\it {Resonant Leptogenesis in the Minimal
  B-L Extended Standard Model at TeV}},  {\em Phys. Rev. D} {\bf 83} (2011)
  093011, [\href{http://arxiv.org/abs/1011.4769}{{\tt arXiv:1011.4769}}].

\bibitem{Okada:2012fs}
N.~Okada, Y.~Orikasa, and T.~Yamada, {\it {Minimal Flavor Violation in the
  Minimal $U(1)_{B-L}$ Model and Resonant Leptogenesis}},  {\em Phys. Rev. D}
  {\bf 86} (2012) 076003, [\href{http://arxiv.org/abs/1207.1510}{{\tt
  arXiv:1207.1510}}].

\bibitem{Davidson:2002qv}
S.~Davidson and A.~Ibarra, {\it {A Lower bound on the right-handed neutrino
  mass from leptogenesis}},  {\em Phys. Lett. B} {\bf 535} (2002) 25--32,
  [\href{http://arxiv.org/abs/hep-ph/0202239}{{\tt hep-ph/0202239}}].

\bibitem{Pilaftsis:2003gt}
A.~Pilaftsis and T.~E.~J. Underwood, {\it {Resonant leptogenesis}},  {\em Nucl.
  Phys. B} {\bf 692} (2004) 303--345,
  [\href{http://arxiv.org/abs/hep-ph/0309342}{{\tt hep-ph/0309342}}].

\bibitem{Newton:2020cog}
O.~Newton, M.~Leo, M.~Cautun, A.~Jenkins, C.~S. Frenk, M.~R. Lovell, J.~C.
  Helly, A.~J. Benson, and S.~Cole, {\it {Constraints on the properties of warm
  dark matter using the satellite galaxies of the Milky Way}},  {\em JCAP} {\bf
  08} (2021) 062, [\href{http://arxiv.org/abs/2011.08865}{{\tt
  arXiv:2011.08865}}].

\bibitem{Banik:2019smi}
N.~Banik, J.~Bovy, G.~Bertone, D.~Erkal, and T.~J.~L. de~Boer, {\it {Novel
  constraints on the particle nature of dark matter from stellar streams}},
  {\em JCAP} {\bf 10} (2021) 043, [\href{http://arxiv.org/abs/1911.02663}{{\tt
  arXiv:1911.02663}}].

\bibitem{KATRIN:2019yun}
{\bf KATRIN} Collaboration, M.~Aker et~al., {\it {Improved Upper Limit on the
  Neutrino Mass from a Direct Kinematic Method by KATRIN}},  {\em Phys. Rev.
  Lett.} {\bf 123} (2019), no.~22 221802,
  [\href{http://arxiv.org/abs/1909.06048}{{\tt arXiv:1909.06048}}].

\bibitem{Dolinski:2019nrj}
M.~J. Dolinski, A.~W.~P. Poon, and W.~Rodejohann, {\it {Neutrinoless
  Double-Beta Decay: Status and Prospects}},  {\em Ann. Rev. Nucl. Part. Sci.}
  {\bf 69} (2019) 219--251, [\href{http://arxiv.org/abs/1902.04097}{{\tt
  arXiv:1902.04097}}].

\end{thebibliography}
\providecommand{\href}[2]{#2}\begingroup\raggedright\endgroup

\end{document}